\documentclass[aps,prl,twocolumn,amsmath,amssymb,nofootinbib,superscriptaddress]{revtex4-2}

\usepackage{graphicx}
\usepackage{bm}
\usepackage[colorlinks]{hyperref}
\usepackage[usenames,svgnames]{xcolor}
\usepackage{makecell}
\usepackage[normalem]{ulem} 
\usepackage{soul}

\newcommand{\loss}{\mathcal{L}}
\newcommand{\trace}{{\rm tr}}
\newcommand{\Hop}{\hat{H}}
\newcommand{\sgp}{\hat{\sigma}^+}
\newcommand{\sgm}{\hat{\sigma}^-}
\newcommand{\sgz}{\hat{\sigma}^z}
\newcommand{\imbalance}{\Delta h}

\newcommand{\Tr}{{\rm Tr}}
\newcommand{\rhoop}{\hat{\rho}}
\newcommand{\Uop}{\hat{U}}
\newcommand{\Gop}{\hat{G}}
\newcommand{\Mop}{\hat{M}}

\newcommand{\Jeff}{\gamma_{{\rm eff}}}
\newcommand{\Ktrain}{K_{{\rm train}}}
\newcommand{\Kval}{K_{{\rm val}}}
\newcommand{\Kpred}{K_{{\rm pred}}}

\newcommand{\EqDef}{\stackrel{\mathrm{def}}{=}}

\newcommand{\hnu}{Key Laboratory of Low-Dimensional Quantum Structures and Quantum Control of Ministry of Education, Department of Physics and Synergetic Innovation Center for Quantum Effects and Applications, Hunan Normal University, Changsha 410081, China
}

\newcommand{\nudt}{Institute for Quantum Information \& State Key Laboratory of High Performance Computing, College of Computer Science and Technology, National University of Defense Technology, Changsha 410073, China
}

\newcommand{\cqt}{Centre for Quantum Technologies, National University of Singapore 117543, Singapore} 
\newcommand{\majulab}{MajuLab, CNRS-UNS-NUS-NTU International Joint Research Unit, UMI 3654, Singapore}

\newcommand{\cas}{Institute of Physics, Chinese Academy of Sciences, Beijing 100190, China
}

\begin{document}

\title{Randomised benchmarking for characterizing and forecasting correlated processes}

\author{Xinfang Zhang}
\thanks{These two authors contributed equally}
\affiliation{\nudt}

\author{Zhihao Wu}
\thanks{These two authors contributed equally}
\affiliation{\nudt}

\author{Gregory A. L. White}
\affiliation{School of Physics and Astronomy, Monash University, Victoria 3800, Australia}
\affiliation{Dahlem Center for Complex Quantum Systems, Freie Universit\"at Berlin, 14195 Berlin, Germany}

\author{Zhongcheng Xiang}
\affiliation{\cas}

\author{Shun Hu}
\affiliation{\nudt}

\author{Zhihui Peng}
\affiliation{\hnu}

\author{Yong Liu}
\affiliation{\nudt}

\author{Dongning Zheng}
\affiliation{\cas}

\author{Xiang Fu}
\affiliation{\nudt}

\author{Anqi Huang}
\affiliation{\nudt}

\author{Dario Poletti} 
\email{dario\_poletti@sutd.edu.sg}
\affiliation{Science, Mathematics and Technology Cluster and Engineering Product Development Pillar, Singapore University of Technology and Design, 8 Somapah Road, 487372 Singapore} 
\affiliation{\cqt} 
\affiliation{\majulab}

\author{Kavan Modi}
\email{kavan.modi@monash.edu}
\affiliation{School of Physics and Astronomy, Monash University, Victoria 3800, Australia}
\affiliation{Quantum for NSW, Sydney 2000, Australia}

\author{Junjie Wu}
\affiliation{\nudt}

\author{Mingtang Deng}
\email{mtdeng@nudt.edu.cn}
\affiliation{\nudt}

\author{Chu Guo}
\email{guochu604b@gmail.com}
\affiliation{\hnu}



\begin{abstract}
The development of fault-tolerant quantum processors relies on the ability to control noise. A particularly insidious form of noise is temporally correlated or non-Markovian noise. By combining randomized benchmarking with supervised machine learning algorithms, we develop a method to learn the details of temporally correlated noise. In particular, we can learn the time-independent evolution operator of system plus bath and this leads to (i) the ability to characterize the degree of non-Markovianity of the dynamics and (ii) the ability to predict the dynamics of the system even beyond the times we have used to train our model. We exemplify this by implementing our method on a superconducting quantum processor. Our experimental results show a drastic change between the Markovian and non-Markovian regimes for the learning accuracies.
\end{abstract}

\maketitle


A major challenge in building near-term quantum computers is noise~\cite{ZhaoPan2022,KrinnerWallraff2022,Google2023}. 
As the number of qubits and the depths of quantum circuits scale up, the fidelity of the output quantum state decreases rapidly, restricting current experiments to low depths~\cite{AruteMartinis2019,WuPan2021,ZhuPan2022}, or few qubits~\cite{Google2020a}.
A quantum device can be affected by both spatially and temporally correlated noise.
Correlated noise can be more harmful then uncorrelated one for scalable quantum error mitigation~\cite{EndoYing2018,EndoYuan2021}, and can lower the threshold of error correcting codes~\cite{NickersonBrown2019,MaskaraJochym2019}. 
Efficient tools to characterize correlated noise are essential for the development of scalable quantum computing technologies.

A variety of techniques have been developed for characterizing Markovian noises, including spatially correlated ones, such as quantum process tomography (QPT)~\cite{ChuangNielsen1997, ArianoPresti2001}, gate set tomography~\cite{KohoutMaunz2017, NielsenKohout2021} and randomized benchmarking (RB)~\cite{EmersonZyczkowski2005,LeviCory2007,KnillWineland2008,MagesanEmerson2011,MagesanEmerson2012}. RB is a highly economical method that consists of averaging over random sequences of gates to estimate the error rates within a given device.

The process tensor framework was recently introduced to expand the applicability of the above tools to time-correlated or non-Markovian noise~\cite{CostaShrapnel2016,PollockModi2018a}. A process tensor is a complete positive mapping from any sequence of $k$ quantum operations to a final state of the system. 
Similar to QPT, the process tensor can be systematically reconstructed with process tensor tomography (PTT), where one applies all the possible sequences of linearly independent quantum operations and performs quantum state tomography~\cite{WhiteModi2020}. Unlike RB, PTT yields detailed information about the noise, which can be used to improve the performance of the quantum device~\cite{WhiteHill2022}.
However, the detailed characterization of complex noise is far more expensive than RB. PTT requires a number of measurements that grows exponentially with $k$. Meanwhile, it is possible to get around this exponential scaling by exploiting the process tensor's natural form as a matrix product density operator with a finite bond dimension~\cite{PollockModi2018a, white2021many,LuchnikovFilippov2019}, which motivates efficient heuristic PTT schemes based on one-dimensional tensor network states~\cite{GuoPoletti2018, GuoPoletti2020,LuchnikovFilippov2020}. Alternatively, one may apply recently-developed shadow-based schemes~\cite{PhysRevLett.130.160401}.

\begin{figure*}
\includegraphics[width=0.9\textwidth]{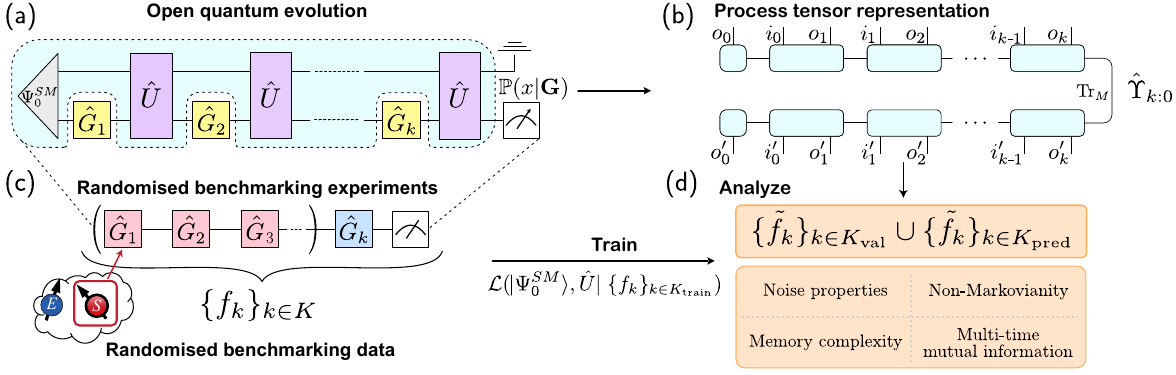}
\caption{(a) The open quantum evolution model where the (non-Markovian) system ($S$) dynamics is induced by coupling to an (unknown) memory $M$ under the $SM$ unitary evolution $\Uop$. $\vert \Psi_0^{SM}\rangle$ denotes the $SM$ initial pure state. A set of quantum operations, denoted as $\Gop_1$ to $\Gop_k$, are performed on $S$ at discrete times $t_1$ to $t_k$. 
(b) The process tensor that encodes all the observable information of system dynamics, which is naturally a matrix product density operator. The indices $i_j, o_{j+1}$ label the input and output system states of $\Uop_{j+1:j}$. 
(c) The standard randomized benchmarking protocol in one-to-one correspondence with (a), with $\Gop_k$ the undo gate. The bottom left cloud shows our experiment setup of two coupled qubits, one as system and the other as environment.
(d) Based on the RB data, our supervised learning algorithm reconstructs the hidden OQE model, which, in combination with its process tensor representation, allows us to predict the (future) system dynamics on the two testing datasets $\Kval$ and $\Kpred$, as well as to analyze the noise properties underlying the system dynamics.}   \label{fig:fig1} 
\end{figure*}

In this work, we develop a method to reconstruct the process tensors by applying supervised machine learning methods to RB data. Our scheme thus inherits the simplicity of RB while capturing the complexity of multi-time non-Markovian system dynamics. It enables one to characterize the process, including concrete measures of non-Markovianity, while its computational cost is related to the memory size of the process, which is often small for real quantum hardware~\cite{WhiteHill2022}. 

As a proof of principle, we demonstrate our scheme on a superconducting quantum processor, where we couple a ``system'' qubit to an ``environment'' qubit with tunable coupling strength, going from weak to strong, resulting in system dynamics from nearly Markovian to highly non-Markovian. We observe a sharp change in learning accuracy between these two regimes; very high learning accuracy in the Markovian regime, and generally lower accuracy in the highly non-Markovian regime, although this can be systematically improved by using larger memory models. In both cases, we can predict future dynamics beyond the times used for training.

\textit{Open quantum evolution model and the process tensor framework.} 
Stationary classical non-Markovian processes can always be written as hidden Markov models~\cite{CrutchfieldYong1989, ShaliziCrutchfield2001}.
Similarly, for quantum processes with time-independent noise, one could reconstruct a quantum version of the hidden Markov model, referred to as the open quantum evolution (OQE) model, which describes the overall unitary dynamics of the system coupled to a minimal environment~\cite{Guo2022a}, which we call \textit{memory} $M$. Once obtained, the OQE model contains all the information of the system dynamics, which can be used to compute process tensors of any steps and predict all the future dynamics.

Without loss of generality, we assume a pure system-memory ($SM$) initial state $\vert \Psi_0^{SM}\rangle$.
We consider the discretized $SM$ dynamics from time steps $t_1$ to $t_k$ ($t_0=0$), and denote the unitary $SM$ evolutionary operator from  $t_{j-1}$ to $t_j$ as $\Uop_{j,j-1}$. At each step $j$ we apply a quantum operation $\Gop_j$ (unitary operation or measurement) on $S$. The overall $SM$ dynamics can be written as
\begin{align}\label{eq:oqe}
\vert \Psi_k^{SM}\rangle = \Uop_{k+1:k} \Gop_k \cdots \Gop_2 \Uop_{2:1} \Gop_1 \Uop_{1:0}\vert \Psi_0^{SM}\rangle,
\end{align}
which is shown in Fig.~\ref{fig:fig1}(a).
Eq.~\eqref{eq:oqe} also defines the process tensor $\hat{\Upsilon}_{k:0}$ as a mapping from initial state of the system $\rhoop^S_0$, together with $\{\Gop_i\}_{i=1}^k \EqDef \{\Gop_1, \dots, \Gop_k\}$, into the final state $\rhoop^S_k=\trace_M(\vert \Psi_k^{SM}\rangle \langle \Psi_k^{SM}\vert )$, as shown in Fig.~\ref{fig:fig1}(b) (See Supplementary for detailed construction of $\hat{\Upsilon}_{k:0}$ from OQE~\cite{supp}). 
$\hat{\Upsilon}_{k:0}$ contains all the observable information of the $k$-step system dynamics and is uniquely defined (in contrast OQE is not unique~\cite{Guo2022a}).
In the next, we focus on time-independent noise with $\Uop_{j:j-1}=\Uop$ for any $j$. In this case, the open quantum dynamics is completely determined by $\vert \Psi_0^{SM}\rangle$ and $\Uop$, and our goal is to determine them by performing RB on the system only.

\textit{Reconstructing the OQE model with RB.} 
For RB one first randomly generates $n$ sequences, denoted as $\{ \{\Gop^1_i\}_{i=1}^k, \dots, \{\Gop^n_i\}_{i=1}^k \}$, where in each sequence the last operation $\Gop_k^l$ is understood as the undo gate $\Gop_k^l = (\Gop_{k-1}^l \cdots \Gop_2^l \Gop_1^l)^{\dagger}$ used to isolate the noise effects. 
Then the RB protocol proceeds as follows: (1) Preparing an initial state $\rhoop^S_0$ of the system; (2) Applying each $\{\Gop^l_i\}_{i=1}^k$ onto $\rhoop^S_0$ to obtain the final state $\rhoop^{S, l}_k$ and measuring $f^l_k = \Tr(\Mop\rhoop^{S, l}_k)$ for a positive operator-valued measure element $\Mop$; (3) Computing the average $F_k = \sum_{l=1}^n f_k^l / n$ and repeating (1) and (2) for different $k$. For Markovian noise that is also gate-independent, $F_k$ can be well approximated by an exponential decay $F_k = ap^k + b$ where $a$, $b$ and $p$ are constants~\cite{MagesanEmerson2011,MagesanEmerson2012}. For non-Markovian noise, instead, non-exponential behavior of $F_k$ is expected in general~\cite{RomeroHsieh2021}. 
The RB protocol can be naturally understood in the process tensor framework, which is demonstrated in Fig.~\ref{fig:fig1}(c) in correspondence with Fig.~\ref{fig:fig1}(a). 

Based on the RB data, we propose a variational scheme to reconstruct the hidden OQE model by minimizing the mean square loss between the predicted outcomes $\tilde{f}^l_k = \langle \Psi^{SM, l}_k \vert \Mop \vert \Psi^{SM, l}_k \rangle $ and the experiment outcomes $f^l_k$:
\begin{align}\label{eq:loss}
\loss(\vert \Psi_0^{SM}\rangle, \Uop) = \frac{1}{n |\Ktrain|} \sum_{k \in \Ktrain} \sum_{l=1}^n \left(f^l_k - \tilde{f}^l_k \right)^2,
\end{align}
where $\Ktrain$ is the set of $k$ used for training. We use the BFGS optimizer~\cite{BFGS},
with $\Uop$ parameterized as in Ref.~\cite{ReckBertani1994} and randomly initialized with a predefined memory size $\chi$. The gradient is computed by automatic differentiation~\cite{GuoPoletti2021}.
Once an optimal OQE model has been obtained, one can use it to predict the output quantum state $\rhoop^S_{k'}$ of the system for any sequence $\{\Gop_i\}_{i=1}^{k'}$ ($k'$ may not be in $\Ktrain$) as shown in Fig.~\ref{fig:fig1}(d). 

\textit{Experimental setup.} To demonstrate our method, we apply it to reconstruct temporally correlated noise on a superconducting quantum processor. We use two capacitive-coupled transmon qubits, one as system ($S$) and the other as environment ($E$). Note that we have differentiated between the physical environment $E$ of the quantum processor and memory $M$ that goes into the OQE, where $M$ includes the qubit $E$ and can include other effects that we cannot directly control. The $SE$ Hamiltonian is
$\Hop = J(\sgp_S \sgm_E + \sgp_E \sgm_S ) + h_S\sgz_S + h_E\sgz_E$,
where $J$ is the coupling strength, $h_{S/E}$ is the local energy for the system or environment qubit. 
Once again, we highlight the fact that a one qubit memory is sufficient for modelling realistic quantum computers~\cite{WhiteModi2020,white2021many,WhiteHill2022}. 

\begin{figure}
\includegraphics[width=\columnwidth]{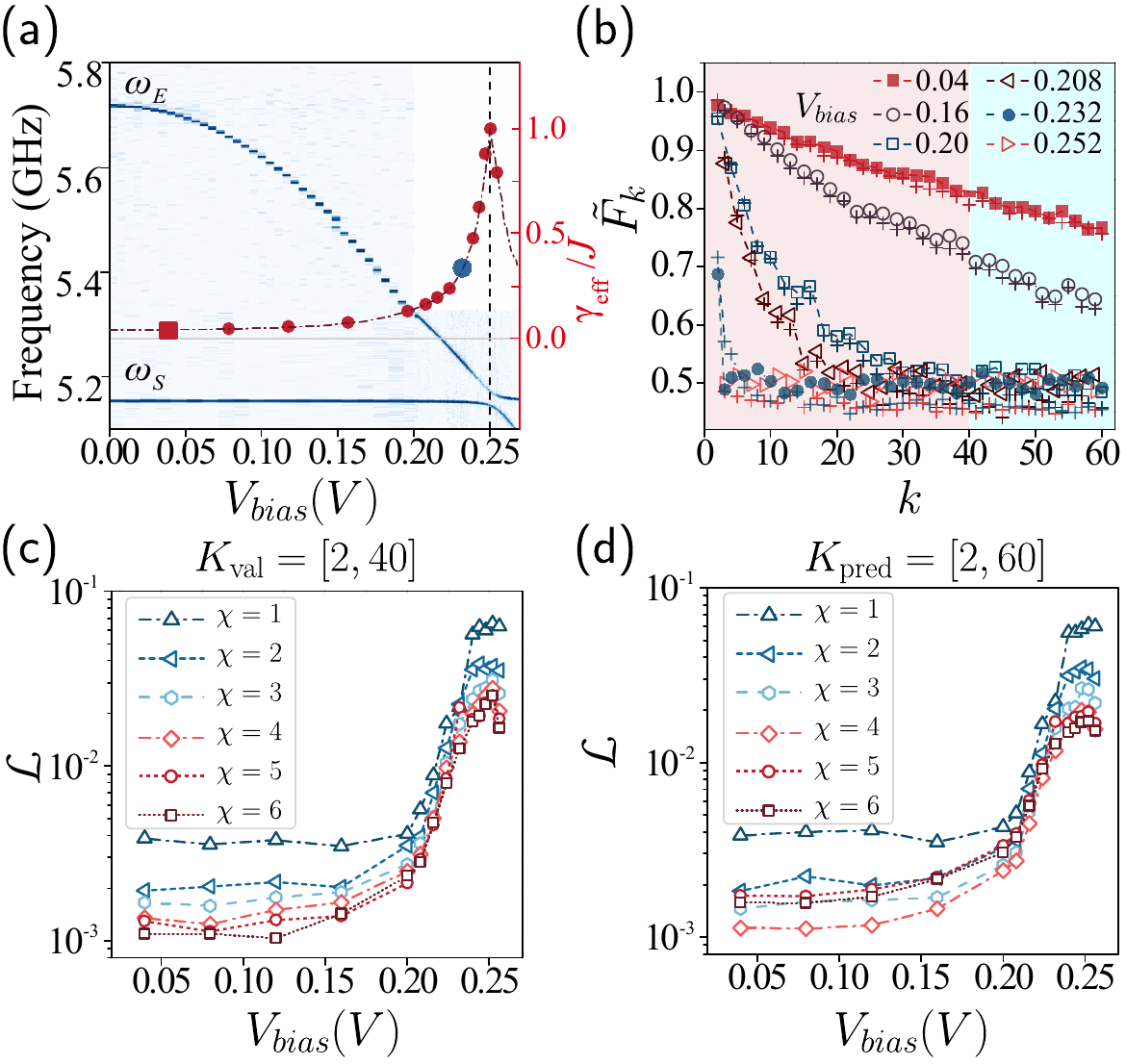}
\caption{
(a) 
The frequencies $\omega_{S/E}=h_{S/E}/\hbar$ of the system and environment qubits (left axis) and the effective coupling strength $\Jeff$ (right axis) as functions of $V_{bias}$.
(b) The averaged measurement outcome as a function of $k$ for selected values of $V_{bias}$, on testing dataset $\Kpred$ that goes beyond the training times.
The markers in the legend represent the predicted $\tilde{F}_k$ from the reconstructed OQE with $\chi=5$. The dashed lines with the same colors and marked with $+$ are the corresponding experiment outcomes.
The two solid markers correspond to the two points in (a) with the same symbols.
(c,d) The loss values $\loss$ as a function $V_{bias}$, evaluated on (c) $\Kval$ and (d) $\Kpred$ respectively, where the results for different $\chi$s used for constructing the hidden OQE models are shown. 
}   \label{fig:fig2} 
\end{figure}

We apply the standard RB protocol on $S$ as depicted in Fig.~\ref{fig:fig1}(c). The initial state is chosen as $\rhoop^S_0 = \vert 0^S\rangle\langle 0^S\vert$. 
While it is not necessary for our algorithm, for a more straightforward implementation we consider that the $SM$ initial state is separable: $\vert \Psi_0^{SM}\rangle = \vert 0^S\rangle \otimes \vert \Psi_0^M\rangle$, and we can simplify Eq.~\eqref{eq:loss} by fixing $\vert \Psi_0^M\rangle = \vert 0^M\rangle$ without loss of generality (one can change the environment basis without any observable effects). 
As a result, only $\Uop$ remains to be determined.
To tune the system dynamics from Markovian to non-Markovian, one needs to tune the coupling between $S$ and $M$.
In our experiment, $J$ is kept as a constant, but we tune the effective coupling strength $\Jeff = 2J^2/\imbalance$ by changing the imbalance $\imbalance = h_S - h_E$ via the voltage bias $V_{bias}$, as shown in Fig.~\ref{fig:fig2}(a).



\textit{The reconstruction accuracy.} 
In our experiment, each gate operation takes about $20 \times i$ ns, with $i = 1, 2, 3$ depending on the number of native gates obtained through Epstein decomposition~\cite{EpsteinGambetta2014}.
The idle time between gates is set to be $100$ ns. In this way, the duration between successive time steps in OQE could be slightly different, which would break our assumption of time-independent $\Uop$. Nevertheless, our results later show that our reconstructed models are still accurate.
For each $V_{bias}$ in Fig.~\ref{fig:fig2}(a), we independently prepare two datasets with $k \in [2,40]$ and $k \in [2,60]$ respectively, and for each $k$ we prepare $n=200$ data pairs.
We take $60\%$ (for each $k$) of the first dataset for training ($\Ktrain$), and the rest of the first dataset (denoted as $\Kval$ instead) as well as the whole second dataset ($\Kpred$) for testing.
For training, we ramp up $\chi$ (dimension of $M$) from $1$ to $6$ such that our model becomes increasingly more expressive. For each $\chi$, we use the BFGS optimizer with at most $200$ iterations to find an optimal $\Uop$ as a $2\chi\times 2\chi$ unitary matrix. We run the optimization for each instance for $5$ times and choose the one with the lowest loss value as our final result. 

In Fig.~\ref{fig:fig2}(c,d), we show the loss values for the two testing datasets $\Kval$ and $\Kpred$ respectively.
In both cases, we can see that for $V_{bias} \leq 0.2$ we can obtain very low loss with a small $\chi$, while for $V_{bias}>0.2$ one needs larger $\chi$ to reach lower loss. Importantly, the model trained for $\Ktrain=[2, 40]$ can be well generalized to $\Kpred=[2, 60]$, which shows that our method is capable of predicting future dynamics. In addition, the OQE trained with $\chi=4$ works better for $\Kpred$ than $\chi=5,6$, especially for $V_{bias} < 0.22$, which is a sign of overfitting in the near Markovian regime for large $\chi$.

To better visualize the power of our trained OQE model, in Fig.~\ref{fig:fig2}(b) we directly plot the average predicted measurement outcomes $\tilde{F}_k = \sum_{l=1}^n\tilde{f}_k^l / n$ from the OQE reconstructed with $\chi=5$, compared to $F_k$ obtained from experiment for $\Kpred$. We can see that in the near-Markovian regime with $V_{bias}=0.04,0.16$, there is a very good matching between them, while in the non-Markovian regime, the discrepancy becomes larger, especially for larger $k$. There is a constant bias between the predicted values and the experimental results for large $k$, which indicates a measurement bias from the experiment that has not been taken into account in our method.


\textit{Quantifying the properties of multi-time processes.}
The process tensor, obtained from the OQE model, is a Hermitian, positive, unit-trace multipartite matrix. In other words, it is a multi-time density matrix whose correlations quantify non-Markovianity. Here, we consider three different properties of the process tensor: 1) its entropy; 2) its multipartite non-Markovianity; and 3) non-Markovianity across two times. As such, we need to define von Neumann entropy $\mathcal{S}(x) := -\mbox{tr}[\rho_x\log(\rho_x)]$ and mutual information $\mathcal{I}(x,y) := \mathcal{S}(x) + \mathcal{S}(y) - \mathcal{S}(xy)$.

\begin{figure}
\includegraphics[width=\columnwidth]{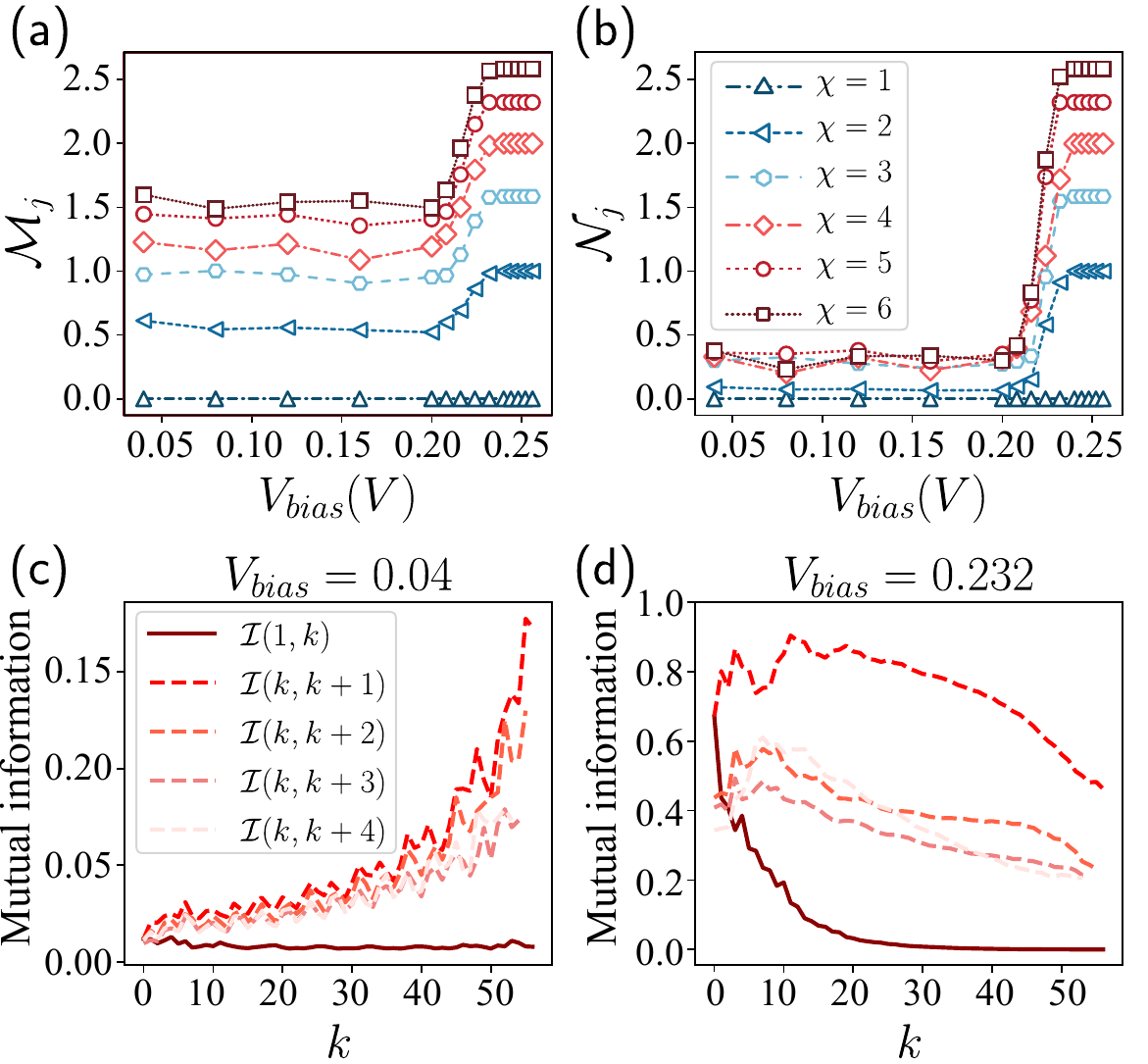}
\caption{(a) The memory complexity $\mathcal{M}_j$ and (b) the non-Markovianity $\mathcal{N}_j$ evaluated at $j=40$ based on the process tensor representation of the reconstructed OQE, for $\chi$ increased from $1$ to $6$. (c, d) Mutual information calculated for two specific values of $V_{bias}$, corresponding to the two special markers in Fig.~\ref{fig:fig2}(a).}  \label{fig:fig5} 
\end{figure}


For the first measure, we compute the entropy of the whole process tensor $\mathcal{M}_j = \mathcal{S}(\hat{\Upsilon}_{j:0})$, which quantifies the level of noise in the process. For OQE, $\mathcal{M}_j$ is the same as the final entropy of the $M$ space, thus it is sometimes referred to as \textit{memory complexity}~\cite{Guo2022a} and it resembles statistical complexity of classical stochastic process~\cite{YangGu2018}.
The second measure quantifies the multi-time correlation between the past and the future of the process. To do so, we first vectorise the process tensor, i.e.,  $\hat{\Upsilon}_{k:0} \to \mbox{vec}(\hat{\Upsilon}_{k:0})/ \|\hat{\Upsilon}_{k:0}\|_2$, with $\|\hat{\Upsilon}_{k:0}\|_2$ the normalisation. The entropy of a subpart $x$ of this pure state vanishes iff the process is Markovian~\cite{Guo2022c}. We take $x$ to be times $0$ to $j$ and denote the entropy as $\mathcal{N}_j$, which captures the correlations between the past and the future.
Finally, we compute the mutual information between marginals of the process tensor; from $\hat\Upsilon_{k:0}$, we obtain a bipartite process tensor $\hat{\Upsilon}_{x,y}$ by contracting all $o_j$ with $i_j$, i.e., inserting the identity gates at all time slots $j$. Indices $o_{x-1},o_{y-1}$ are traced out and a $\left|0\right>$ is inserted at $i_x$. The mutual information of $\hat{\Upsilon}_{x,y}$ quantifies four time correlation which vanishes for all Markovian processes. 

The three quantities above capture different aspects of the non-Markovian process and are plotted in Fig.~\ref{fig:fig5}. Panels (a,b) display $\mathcal{M}_j$ and $\mathcal{N}_j$, respectively to show a sharp transition from the Markovian regime for $V_{bias}\leq 0.2$ to the non-Markovian regime for $V_{bias}> 0.2$.
Neither measure converges with $\chi$ for $V_{bias} > 0.23$, which means that a larger $\chi$ (and more training data) may be required for the hidden OQE. Interestingly, for $V_{bias}\leq 0.2$, $\mathcal{M}_j$ is much larger than $\mathcal{N}_j$. This could mean that the underlying quantum dynamics can be well approximated by a Markovian but non-unitary process. 
To further examine this, we compute the mutual information for $V_{bias}=0.04$ and $V_{bias}=0.232$ respectively in panels (c,d). 
In the first case, the mutual information is small and gradually increases with $k$, indicating an evolution of short-range memory with the dynamics. Owing to the lower coupling, more time is needed to generate the required interaction to accumulate temporal correlations. Moreover, as one would expect, these correlations reduce with temporal separation $\Delta k$. Nevertheless, one can see that even in a low-coupling regime a coherent system can eventually develop non-Markovian features.
Meanwhile, in the second case, the mutual information starts very high and quickly saturates (note the y-axis scale difference between c/d). Because we have fixed interaction with a small environment, the quantity $\mathcal{I}(1,k)$ decays exponentially. But the closer-in-time correlations maintain a large steady state with $k$ until they are slowly reduced by dissipation. Using our tools, we see that the dynamics of memory can be studied in company with the usual information provided by RB curves.



\textit{Summary.} We have proposed an experimentally friendly scheme to characterize temporally correlated noise in open quantum dynamics based on data from randomized benchmarking experiments. 
We demonstrated our method on a superconducting quantum processor, where we tune the quantum dynamic of the system qubit from Markovian to highly non-Markovian by tuning the effective coupling strength to an environment qubit. Our results show that, close to the Markovian regime, we can reconstruct an OQE model with a very small memory size and with high prediction accuracy on the testing datasets. In the highly non-Markovian regime, the reconstruction accuracy is generally lower but can be systematically improved by using a larger memory size. 
In both cases the reconstructed OQE model can well predict the observed and even unobserved system dynamics.
We computed three different measures of non-Markovianity using the process tensor obtained from reconstructed OQE and we find that they indeed have a close correspondence with the Markovian or non-Markovian behaviors of the system dynamics. Our method thus opens up the possibility of quantifying temporally correlated noise in quantum devices based on existing RB data.

\begin{acknowledgements}

This work was supported by the Open Research Fund from State Key Laboratory of High Performance Computing of China (Grant No. 202201-00), the Hunan Provincial Science Fund for Distinguished Young Scholars (Grant No. 2021JJ10043), and the Innovation Program for Quantum Science and Technology (Grant No. 2021ZD030240). D.P. acknowledges support from the National Research Foundation, Singapore under its
QEP2.0 programme (NRF2021-QEP2-02-P03). 

\end{acknowledgements}

\bibliographystyle{apsrev4-1}
\bibliography{refs_v1}

\begin{thebibliography}{41}%
\makeatletter
\providecommand \@ifxundefined [1]{%
 \@ifx{#1\undefined}
}%
\providecommand \@ifnum [1]{%
 \ifnum #1\expandafter \@firstoftwo
 \else \expandafter \@secondoftwo
 \fi
}%
\providecommand \@ifx [1]{%
 \ifx #1\expandafter \@firstoftwo
 \else \expandafter \@secondoftwo
 \fi
}%
\providecommand \natexlab [1]{#1}%
\providecommand \enquote  [1]{``#1''}%
\providecommand \bibnamefont  [1]{#1}%
\providecommand \bibfnamefont [1]{#1}%
\providecommand \citenamefont [1]{#1}%
\providecommand \href@noop [0]{\@secondoftwo}%
\providecommand \href [0]{\begingroup \@sanitize@url \@href}%
\providecommand \@href[1]{\@@startlink{#1}\@@href}%
\providecommand \@@href[1]{\endgroup#1\@@endlink}%
\providecommand \@sanitize@url [0]{\catcode `\\12\catcode `\$12\catcode
  `\&12\catcode `\#12\catcode `\^12\catcode `\_12\catcode `\%12\relax}%
\providecommand \@@startlink[1]{}%
\providecommand \@@endlink[0]{}%
\providecommand \url  [0]{\begingroup\@sanitize@url \@url }%
\providecommand \@url [1]{\endgroup\@href {#1}{\urlprefix }}%
\providecommand \urlprefix  [0]{URL }%
\providecommand \Eprint [0]{\href }%
\providecommand \doibase [0]{http://dx.doi.org/}%
\providecommand \selectlanguage [0]{\@gobble}%
\providecommand \bibinfo  [0]{\@secondoftwo}%
\providecommand \bibfield  [0]{\@secondoftwo}%
\providecommand \translation [1]{[#1]}%
\providecommand \BibitemOpen [0]{}%
\providecommand \bibitemStop [0]{}%
\providecommand \bibitemNoStop [0]{.\EOS\space}%
\providecommand \EOS [0]{\spacefactor3000\relax}%
\providecommand \BibitemShut  [1]{\csname bibitem#1\endcsname}%
\let\auto@bib@innerbib\@empty
\bibitem [{\citenamefont {Zhao}\ \emph {et~al.}(2022)\citenamefont {Zhao},
  \citenamefont {Ye}, \citenamefont {Huang}, \citenamefont {Zhang},
  \citenamefont {Wu}, \citenamefont {Guan}, \citenamefont {Zhu}, \citenamefont
  {Wei}, \citenamefont {He},\ and\ \citenamefont {Cao}}]{ZhaoPan2022}%
  \BibitemOpen
  \bibfield  {author} {\bibinfo {author} {\bibfnamefont {Y.}~\bibnamefont
  {Zhao}}, \bibinfo {author} {\bibfnamefont {Y.}~\bibnamefont {Ye}}, \bibinfo
  {author} {\bibfnamefont {H.-L.}\ \bibnamefont {Huang}}, \bibinfo {author}
  {\bibfnamefont {Y.}~\bibnamefont {Zhang}}, \bibinfo {author} {\bibfnamefont
  {D.}~\bibnamefont {Wu}}, \bibinfo {author} {\bibfnamefont {H.}~\bibnamefont
  {Guan}}, \bibinfo {author} {\bibfnamefont {Q.}~\bibnamefont {Zhu}}, \bibinfo
  {author} {\bibfnamefont {Z.}~\bibnamefont {Wei}}, \bibinfo {author}
  {\bibfnamefont {T.}~\bibnamefont {He}}, \ and\ \bibinfo {author}
  {\bibfnamefont {S.~o.}\ \bibnamefont {Cao}},\ }\href {\doibase
  10.1103/PhysRevLett.129.030501} {\bibfield  {journal} {\bibinfo  {journal}
  {Phys. Rev. Lett.}\ }\textbf {\bibinfo {volume} {129}},\ \bibinfo {pages}
  {030501} (\bibinfo {year} {2022})}\BibitemShut {NoStop}%
\bibitem [{\citenamefont {Krinner}\ \emph {et~al.}(2022)\citenamefont
  {Krinner}, \citenamefont {Lacroix}, \citenamefont {Remm}, \citenamefont
  {Di~Paolo}, \citenamefont {Genois}, \citenamefont {Leroux}, \citenamefont
  {Hellings}, \citenamefont {Lazar}, \citenamefont {Swiadek}, \citenamefont
  {Herrmann} \emph {et~al.}}]{KrinnerWallraff2022}%
  \BibitemOpen
  \bibfield  {author} {\bibinfo {author} {\bibfnamefont {S.}~\bibnamefont
  {Krinner}}, \bibinfo {author} {\bibfnamefont {N.}~\bibnamefont {Lacroix}},
  \bibinfo {author} {\bibfnamefont {A.}~\bibnamefont {Remm}}, \bibinfo {author}
  {\bibfnamefont {A.}~\bibnamefont {Di~Paolo}}, \bibinfo {author}
  {\bibfnamefont {E.}~\bibnamefont {Genois}}, \bibinfo {author} {\bibfnamefont
  {C.}~\bibnamefont {Leroux}}, \bibinfo {author} {\bibfnamefont
  {C.}~\bibnamefont {Hellings}}, \bibinfo {author} {\bibfnamefont
  {S.}~\bibnamefont {Lazar}}, \bibinfo {author} {\bibfnamefont
  {F.}~\bibnamefont {Swiadek}}, \bibinfo {author} {\bibfnamefont
  {J.}~\bibnamefont {Herrmann}},  \emph {et~al.},\ }\href {\doibase
  10.1038/s41586-022-04566-8} {\bibfield  {journal} {\bibinfo  {journal}
  {Nature}\ }\textbf {\bibinfo {volume} {605}},\ \bibinfo {pages} {669}
  (\bibinfo {year} {2022})}\BibitemShut {NoStop}%
\bibitem [{\citenamefont {AI}(2023)}]{Google2023}%
  \BibitemOpen
  \bibfield  {author} {\bibinfo {author} {\bibfnamefont {G.~Q.}\ \bibnamefont
  {AI}},\ }\href {\doibase 10.1038/s41586-022-05434-1} {\bibfield  {journal}
  {\bibinfo  {journal} {Nature}\ }\textbf {\bibinfo {volume} {614}},\ \bibinfo
  {pages} {676} (\bibinfo {year} {2023})}\BibitemShut {NoStop}%
\bibitem [{\citenamefont {Arute}\ \emph {et~al.}(2019)\citenamefont {Arute},
  \citenamefont {Arya}, \citenamefont {Babbush}, \citenamefont {Bacon},
  \citenamefont {Bardin}, \citenamefont {Barends}, \citenamefont {Biswas},
  \citenamefont {Boixo}, \citenamefont {Brandao}, \citenamefont {Buell} \emph
  {et~al.}}]{AruteMartinis2019}%
  \BibitemOpen
  \bibfield  {author} {\bibinfo {author} {\bibfnamefont {F.}~\bibnamefont
  {Arute}}, \bibinfo {author} {\bibfnamefont {K.}~\bibnamefont {Arya}},
  \bibinfo {author} {\bibfnamefont {R.}~\bibnamefont {Babbush}}, \bibinfo
  {author} {\bibfnamefont {D.}~\bibnamefont {Bacon}}, \bibinfo {author}
  {\bibfnamefont {J.~C.}\ \bibnamefont {Bardin}}, \bibinfo {author}
  {\bibfnamefont {R.}~\bibnamefont {Barends}}, \bibinfo {author} {\bibfnamefont
  {R.}~\bibnamefont {Biswas}}, \bibinfo {author} {\bibfnamefont
  {S.}~\bibnamefont {Boixo}}, \bibinfo {author} {\bibfnamefont {F.~G. S.~L.}\
  \bibnamefont {Brandao}}, \bibinfo {author} {\bibnamefont {Buell}},  \emph
  {et~al.},\ }\href {\doibase doi.org/10.1038/s41586-019-1666-5} {\bibfield
  {journal} {\bibinfo  {journal} {Nature}\ }\textbf {\bibinfo {volume} {574}},\
  \bibinfo {pages} {505} (\bibinfo {year} {2019})}\BibitemShut {NoStop}%
\bibitem [{\citenamefont {Wu}\ \emph {et~al.}(2021)\citenamefont {Wu},
  \citenamefont {Bao}, \citenamefont {Cao}, \citenamefont {Chen}, \citenamefont
  {Chen}, \citenamefont {Chen}, \citenamefont {Chung}, \citenamefont {Deng},
  \citenamefont {Du}, \citenamefont {Fan} \emph {et~al.}}]{WuPan2021}%
  \BibitemOpen
  \bibfield  {author} {\bibinfo {author} {\bibfnamefont {Y.}~\bibnamefont
  {Wu}}, \bibinfo {author} {\bibfnamefont {W.-S.}\ \bibnamefont {Bao}},
  \bibinfo {author} {\bibfnamefont {S.}~\bibnamefont {Cao}}, \bibinfo {author}
  {\bibfnamefont {F.}~\bibnamefont {Chen}}, \bibinfo {author} {\bibfnamefont
  {M.-C.}\ \bibnamefont {Chen}}, \bibinfo {author} {\bibfnamefont
  {X.}~\bibnamefont {Chen}}, \bibinfo {author} {\bibfnamefont {T.-H.}\
  \bibnamefont {Chung}}, \bibinfo {author} {\bibfnamefont {H.}~\bibnamefont
  {Deng}}, \bibinfo {author} {\bibfnamefont {Y.}~\bibnamefont {Du}}, \bibinfo
  {author} {\bibfnamefont {D.}~\bibnamefont {Fan}},  \emph {et~al.},\ }\href
  {\doibase 10.1103/PhysRevLett.127.180501} {\bibfield  {journal} {\bibinfo
  {journal} {Phys. Rev. Lett.}\ }\textbf {\bibinfo {volume} {127}},\ \bibinfo
  {pages} {180501} (\bibinfo {year} {2021})}\BibitemShut {NoStop}%
\bibitem [{\citenamefont {Zhu}\ \emph {et~al.}(2022)\citenamefont {Zhu},
  \citenamefont {Cao}, \citenamefont {Chen}, \citenamefont {Chen},
  \citenamefont {Chen} \emph {et~al.}}]{ZhuPan2022}%
  \BibitemOpen
  \bibfield  {author} {\bibinfo {author} {\bibfnamefont {Q.}~\bibnamefont
  {Zhu}}, \bibinfo {author} {\bibfnamefont {S.}~\bibnamefont {Cao}}, \bibinfo
  {author} {\bibfnamefont {F.}~\bibnamefont {Chen}}, \bibinfo {author}
  {\bibfnamefont {M.-C.}\ \bibnamefont {Chen}}, \bibinfo {author}
  {\bibfnamefont {X.}~\bibnamefont {Chen}},  \emph {et~al.},\ }\href {\doibase
  doi.org/10.1016/j.scib.2021.10.017} {\bibfield  {journal} {\bibinfo
  {journal} {Science Bulletin}\ }\textbf {\bibinfo {volume} {67}},\ \bibinfo
  {pages} {240} (\bibinfo {year} {2022})}\BibitemShut {NoStop}%
\bibitem [{\citenamefont {Arute}\ \emph {et~al.}(2020)\citenamefont {Arute},
  \citenamefont {Arya}, \citenamefont {Babbush}, \citenamefont {Bacon},
  \citenamefont {Bardin}, \citenamefont {Barends}, \citenamefont {Boixo},
  \citenamefont {Broughton}, \citenamefont {Buckley} \emph
  {et~al.}}]{Google2020a}%
  \BibitemOpen
  \bibfield  {author} {\bibinfo {author} {\bibfnamefont {F.}~\bibnamefont
  {Arute}}, \bibinfo {author} {\bibfnamefont {K.}~\bibnamefont {Arya}},
  \bibinfo {author} {\bibfnamefont {R.}~\bibnamefont {Babbush}}, \bibinfo
  {author} {\bibfnamefont {D.}~\bibnamefont {Bacon}}, \bibinfo {author}
  {\bibfnamefont {J.~C.}\ \bibnamefont {Bardin}}, \bibinfo {author}
  {\bibfnamefont {R.}~\bibnamefont {Barends}}, \bibinfo {author} {\bibfnamefont
  {S.}~\bibnamefont {Boixo}}, \bibinfo {author} {\bibfnamefont
  {M.}~\bibnamefont {Broughton}}, \bibinfo {author} {\bibfnamefont {B.~B.}\
  \bibnamefont {Buckley}},  \emph {et~al.},\ }\href {\doibase
  10.1126/science.abb9811} {\bibfield  {journal} {\bibinfo  {journal}
  {Science}\ }\textbf {\bibinfo {volume} {369}},\ \bibinfo {pages} {1084}
  (\bibinfo {year} {2020})}\BibitemShut {NoStop}%
\bibitem [{\citenamefont {Endo}\ \emph {et~al.}(2018)\citenamefont {Endo},
  \citenamefont {Benjamin},\ and\ \citenamefont {Li}}]{EndoYing2018}%
  \BibitemOpen
  \bibfield  {author} {\bibinfo {author} {\bibfnamefont {S.}~\bibnamefont
  {Endo}}, \bibinfo {author} {\bibfnamefont {S.~C.}\ \bibnamefont {Benjamin}},
  \ and\ \bibinfo {author} {\bibfnamefont {Y.}~\bibnamefont {Li}},\ }\href
  {\doibase 10.1103/PhysRevX.8.031027} {\bibfield  {journal} {\bibinfo
  {journal} {Phys. Rev. X}\ }\textbf {\bibinfo {volume} {8}},\ \bibinfo {pages}
  {031027} (\bibinfo {year} {2018})}\BibitemShut {NoStop}%
\bibitem [{\citenamefont {Endo}\ \emph {et~al.}(2021)\citenamefont {Endo},
  \citenamefont {Cai}, \citenamefont {Benjamin},\ and\ \citenamefont
  {Yuan}}]{EndoYuan2021}%
  \BibitemOpen
  \bibfield  {author} {\bibinfo {author} {\bibfnamefont {S.}~\bibnamefont
  {Endo}}, \bibinfo {author} {\bibfnamefont {Z.}~\bibnamefont {Cai}}, \bibinfo
  {author} {\bibfnamefont {S.~C.}\ \bibnamefont {Benjamin}}, \ and\ \bibinfo
  {author} {\bibfnamefont {X.}~\bibnamefont {Yuan}},\ }\href {\doibase
  10.7566/JPSJ.90.032001} {\bibfield  {journal} {\bibinfo  {journal} {Journal
  of the Physical Society of Japan}\ }\textbf {\bibinfo {volume} {90}},\
  \bibinfo {pages} {032001} (\bibinfo {year} {2021})}\BibitemShut {NoStop}%
\bibitem [{\citenamefont {Nickerson}\ and\ \citenamefont
  {Brown}(2019)}]{NickersonBrown2019}%
  \BibitemOpen
  \bibfield  {author} {\bibinfo {author} {\bibfnamefont {N.~H.}\ \bibnamefont
  {Nickerson}}\ and\ \bibinfo {author} {\bibfnamefont {B.~J.}\ \bibnamefont
  {Brown}},\ }\href {\doibase 10.22331/q-2019-04-08-131} {\bibfield  {journal}
  {\bibinfo  {journal} {{Quantum}}\ }\textbf {\bibinfo {volume} {3}},\ \bibinfo
  {pages} {131} (\bibinfo {year} {2019})}\BibitemShut {NoStop}%
\bibitem [{\citenamefont {Maskara}\ \emph {et~al.}(2019)\citenamefont
  {Maskara}, \citenamefont {Kubica},\ and\ \citenamefont
  {Jochym-O'Connor}}]{MaskaraJochym2019}%
  \BibitemOpen
  \bibfield  {author} {\bibinfo {author} {\bibfnamefont {N.}~\bibnamefont
  {Maskara}}, \bibinfo {author} {\bibfnamefont {A.}~\bibnamefont {Kubica}}, \
  and\ \bibinfo {author} {\bibfnamefont {T.}~\bibnamefont {Jochym-O'Connor}},\
  }\href {\doibase 10.1103/PhysRevA.99.052351} {\bibfield  {journal} {\bibinfo
  {journal} {Phys. Rev. A}\ }\textbf {\bibinfo {volume} {99}},\ \bibinfo
  {pages} {052351} (\bibinfo {year} {2019})}\BibitemShut {NoStop}%
\bibitem [{\citenamefont {Chuang}\ and\ \citenamefont
  {Nielsen}(1997)}]{ChuangNielsen1997}%
  \BibitemOpen
  \bibfield  {author} {\bibinfo {author} {\bibfnamefont {I.~L.}\ \bibnamefont
  {Chuang}}\ and\ \bibinfo {author} {\bibfnamefont {M.~A.}\ \bibnamefont
  {Nielsen}},\ }\href {\doibase 10.1080/09500349708231894} {\bibfield
  {journal} {\bibinfo  {journal} {Journal of Modern Optics}\ }\textbf {\bibinfo
  {volume} {44}},\ \bibinfo {pages} {2455} (\bibinfo {year}
  {1997})}\BibitemShut {NoStop}%
\bibitem [{\citenamefont {D'Ariano}\ and\ \citenamefont
  {Lo~Presti}(2001)}]{ArianoPresti2001}%
  \BibitemOpen
  \bibfield  {author} {\bibinfo {author} {\bibfnamefont {G.~M.}\ \bibnamefont
  {D'Ariano}}\ and\ \bibinfo {author} {\bibfnamefont {P.}~\bibnamefont
  {Lo~Presti}},\ }\href {\doibase 10.1103/PhysRevLett.86.4195} {\bibfield
  {journal} {\bibinfo  {journal} {Phys. Rev. Lett.}\ }\textbf {\bibinfo
  {volume} {86}},\ \bibinfo {pages} {4195} (\bibinfo {year}
  {2001})}\BibitemShut {NoStop}%
\bibitem [{\citenamefont {Blume-Kohout}\ \emph {et~al.}(2017)\citenamefont
  {Blume-Kohout}, \citenamefont {Gamble}, \citenamefont {Nielsen},
  \citenamefont {Rudinger}, \citenamefont {Mizrahi}, \citenamefont {Fortier},\
  and\ \citenamefont {Maunz}}]{KohoutMaunz2017}%
  \BibitemOpen
  \bibfield  {author} {\bibinfo {author} {\bibfnamefont {R.}~\bibnamefont
  {Blume-Kohout}}, \bibinfo {author} {\bibfnamefont {J.~K.}\ \bibnamefont
  {Gamble}}, \bibinfo {author} {\bibfnamefont {E.}~\bibnamefont {Nielsen}},
  \bibinfo {author} {\bibfnamefont {K.}~\bibnamefont {Rudinger}}, \bibinfo
  {author} {\bibfnamefont {J.}~\bibnamefont {Mizrahi}}, \bibinfo {author}
  {\bibfnamefont {K.}~\bibnamefont {Fortier}}, \ and\ \bibinfo {author}
  {\bibfnamefont {P.}~\bibnamefont {Maunz}},\ }\href {\doibase
  10.1038/ncomms14485} {\bibfield  {journal} {\bibinfo  {journal} {Nature
  Communications}\ }\textbf {\bibinfo {volume} {8}},\ \bibinfo {pages} {14485}
  (\bibinfo {year} {2017})}\BibitemShut {NoStop}%
\bibitem [{\citenamefont {Nielsen}\ \emph {et~al.}(2021)\citenamefont
  {Nielsen}, \citenamefont {Gamble}, \citenamefont {Rudinger}, \citenamefont
  {Scholten}, \citenamefont {Young},\ and\ \citenamefont
  {Blume-Kohout}}]{NielsenKohout2021}%
  \BibitemOpen
  \bibfield  {author} {\bibinfo {author} {\bibfnamefont {E.}~\bibnamefont
  {Nielsen}}, \bibinfo {author} {\bibfnamefont {J.~K.}\ \bibnamefont {Gamble}},
  \bibinfo {author} {\bibfnamefont {K.}~\bibnamefont {Rudinger}}, \bibinfo
  {author} {\bibfnamefont {T.}~\bibnamefont {Scholten}}, \bibinfo {author}
  {\bibfnamefont {K.}~\bibnamefont {Young}}, \ and\ \bibinfo {author}
  {\bibfnamefont {R.}~\bibnamefont {Blume-Kohout}},\ }\href {\doibase
  10.22331/q-2021-10-05-557} {\bibfield  {journal} {\bibinfo  {journal}
  {{Quantum}}\ }\textbf {\bibinfo {volume} {5}},\ \bibinfo {pages} {557}
  (\bibinfo {year} {2021})}\BibitemShut {NoStop}%
\bibitem [{\citenamefont {Emerson}\ \emph {et~al.}(2005)\citenamefont
  {Emerson}, \citenamefont {Alicki},\ and\ \citenamefont
  {{\.{Z}}yczkowski}}]{EmersonZyczkowski2005}%
  \BibitemOpen
  \bibfield  {author} {\bibinfo {author} {\bibfnamefont {J.}~\bibnamefont
  {Emerson}}, \bibinfo {author} {\bibfnamefont {R.}~\bibnamefont {Alicki}}, \
  and\ \bibinfo {author} {\bibfnamefont {K.}~\bibnamefont {{\.{Z}}yczkowski}},\
  }\href {\doibase 10.1088/1464-4266/7/10/021} {\bibfield  {journal} {\bibinfo
  {journal} {Journal of Optics B: Quantum and Semiclassical Optics}\ }\textbf
  {\bibinfo {volume} {7}},\ \bibinfo {pages} {S347} (\bibinfo {year}
  {2005})}\BibitemShut {NoStop}%
\bibitem [{\citenamefont {L\'evi}\ \emph {et~al.}(2007)\citenamefont {L\'evi},
  \citenamefont {L\'opez}, \citenamefont {Emerson},\ and\ \citenamefont
  {Cory}}]{LeviCory2007}%
  \BibitemOpen
  \bibfield  {author} {\bibinfo {author} {\bibfnamefont {B.}~\bibnamefont
  {L\'evi}}, \bibinfo {author} {\bibfnamefont {C.~C.}\ \bibnamefont {L\'opez}},
  \bibinfo {author} {\bibfnamefont {J.}~\bibnamefont {Emerson}}, \ and\
  \bibinfo {author} {\bibfnamefont {D.~G.}\ \bibnamefont {Cory}},\ }\href
  {\doibase 10.1103/PhysRevA.75.022314} {\bibfield  {journal} {\bibinfo
  {journal} {Phys. Rev. A}\ }\textbf {\bibinfo {volume} {75}},\ \bibinfo
  {pages} {022314} (\bibinfo {year} {2007})}\BibitemShut {NoStop}%
\bibitem [{\citenamefont {Knill}\ \emph {et~al.}(2008)\citenamefont {Knill},
  \citenamefont {Leibfried}, \citenamefont {Reichle}, \citenamefont {Britton},
  \citenamefont {Blakestad}, \citenamefont {Jost}, \citenamefont {Langer},
  \citenamefont {Ozeri}, \citenamefont {Seidelin},\ and\ \citenamefont
  {Wineland}}]{KnillWineland2008}%
  \BibitemOpen
  \bibfield  {author} {\bibinfo {author} {\bibfnamefont {E.}~\bibnamefont
  {Knill}}, \bibinfo {author} {\bibfnamefont {D.}~\bibnamefont {Leibfried}},
  \bibinfo {author} {\bibfnamefont {R.}~\bibnamefont {Reichle}}, \bibinfo
  {author} {\bibfnamefont {J.}~\bibnamefont {Britton}}, \bibinfo {author}
  {\bibfnamefont {R.~B.}\ \bibnamefont {Blakestad}}, \bibinfo {author}
  {\bibfnamefont {J.~D.}\ \bibnamefont {Jost}}, \bibinfo {author}
  {\bibfnamefont {C.}~\bibnamefont {Langer}}, \bibinfo {author} {\bibfnamefont
  {R.}~\bibnamefont {Ozeri}}, \bibinfo {author} {\bibfnamefont
  {S.}~\bibnamefont {Seidelin}}, \ and\ \bibinfo {author} {\bibfnamefont
  {D.~J.}\ \bibnamefont {Wineland}},\ }\href {\doibase
  10.1103/PhysRevA.77.012307} {\bibfield  {journal} {\bibinfo  {journal} {Phys.
  Rev. A}\ }\textbf {\bibinfo {volume} {77}},\ \bibinfo {pages} {012307}
  (\bibinfo {year} {2008})}\BibitemShut {NoStop}%
\bibitem [{\citenamefont {Magesan}\ \emph {et~al.}(2011)\citenamefont
  {Magesan}, \citenamefont {Gambetta},\ and\ \citenamefont
  {Emerson}}]{MagesanEmerson2011}%
  \BibitemOpen
  \bibfield  {author} {\bibinfo {author} {\bibfnamefont {E.}~\bibnamefont
  {Magesan}}, \bibinfo {author} {\bibfnamefont {J.~M.}\ \bibnamefont
  {Gambetta}}, \ and\ \bibinfo {author} {\bibfnamefont {J.}~\bibnamefont
  {Emerson}},\ }\href {\doibase 10.1103/PhysRevLett.106.180504} {\bibfield
  {journal} {\bibinfo  {journal} {Phys. Rev. Lett.}\ }\textbf {\bibinfo
  {volume} {106}},\ \bibinfo {pages} {180504} (\bibinfo {year}
  {2011})}\BibitemShut {NoStop}%
\bibitem [{\citenamefont {Magesan}\ \emph {et~al.}(2012)\citenamefont
  {Magesan}, \citenamefont {Gambetta},\ and\ \citenamefont
  {Emerson}}]{MagesanEmerson2012}%
  \BibitemOpen
  \bibfield  {author} {\bibinfo {author} {\bibfnamefont {E.}~\bibnamefont
  {Magesan}}, \bibinfo {author} {\bibfnamefont {J.~M.}\ \bibnamefont
  {Gambetta}}, \ and\ \bibinfo {author} {\bibfnamefont {J.}~\bibnamefont
  {Emerson}},\ }\href {\doibase 10.1103/PhysRevA.85.042311} {\bibfield
  {journal} {\bibinfo  {journal} {Phys. Rev. A}\ }\textbf {\bibinfo {volume}
  {85}},\ \bibinfo {pages} {042311} (\bibinfo {year} {2012})}\BibitemShut
  {NoStop}%
\bibitem [{\citenamefont {Costa}\ and\ \citenamefont
  {Shrapnel}(2016)}]{CostaShrapnel2016}%
  \BibitemOpen
  \bibfield  {author} {\bibinfo {author} {\bibfnamefont {F.}~\bibnamefont
  {Costa}}\ and\ \bibinfo {author} {\bibfnamefont {S.}~\bibnamefont
  {Shrapnel}},\ }\href {\doibase 10.1088/1367-2630/18/6/063032} {\bibfield
  {journal} {\bibinfo  {journal} {New Journal of Physics}\ }\textbf {\bibinfo
  {volume} {18}},\ \bibinfo {pages} {063032} (\bibinfo {year}
  {2016})}\BibitemShut {NoStop}%
\bibitem [{\citenamefont {Pollock}\ \emph {et~al.}(2018)\citenamefont
  {Pollock}, \citenamefont {Rodr\'{\i}guez-Rosario}, \citenamefont
  {Frauenheim}, \citenamefont {Paternostro},\ and\ \citenamefont
  {Modi}}]{PollockModi2018a}%
  \BibitemOpen
  \bibfield  {author} {\bibinfo {author} {\bibfnamefont {F.~A.}\ \bibnamefont
  {Pollock}}, \bibinfo {author} {\bibfnamefont {C.}~\bibnamefont
  {Rodr\'{\i}guez-Rosario}}, \bibinfo {author} {\bibfnamefont {T.}~\bibnamefont
  {Frauenheim}}, \bibinfo {author} {\bibfnamefont {M.}~\bibnamefont
  {Paternostro}}, \ and\ \bibinfo {author} {\bibfnamefont {K.}~\bibnamefont
  {Modi}},\ }\href {\doibase 10.1103/PhysRevA.97.012127} {\bibfield  {journal}
  {\bibinfo  {journal} {Phys. Rev. A}\ }\textbf {\bibinfo {volume} {97}},\
  \bibinfo {pages} {012127} (\bibinfo {year} {2018})}\BibitemShut {NoStop}%
\bibitem [{\citenamefont {White}\ \emph {et~al.}(2020)\citenamefont {White},
  \citenamefont {Hill}, \citenamefont {Pollock}, \citenamefont {Hollenberg},\
  and\ \citenamefont {Modi}}]{WhiteModi2020}%
  \BibitemOpen
  \bibfield  {author} {\bibinfo {author} {\bibfnamefont {G.~A.}\ \bibnamefont
  {White}}, \bibinfo {author} {\bibfnamefont {C.~D.}\ \bibnamefont {Hill}},
  \bibinfo {author} {\bibfnamefont {F.~A.}\ \bibnamefont {Pollock}}, \bibinfo
  {author} {\bibfnamefont {L.~C.}\ \bibnamefont {Hollenberg}}, \ and\ \bibinfo
  {author} {\bibfnamefont {K.}~\bibnamefont {Modi}},\ }\href {\doibase
  doi.org/10.1038/s41467-020-20113-3} {\bibfield  {journal} {\bibinfo
  {journal} {Nature Communications}\ }\textbf {\bibinfo {volume} {11}},\
  \bibinfo {pages} {6301} (\bibinfo {year} {2020})}\BibitemShut {NoStop}%
\bibitem [{\citenamefont {White}\ \emph {et~al.}(2022)\citenamefont {White},
  \citenamefont {Pollock}, \citenamefont {Hollenberg}, \citenamefont {Modi},\
  and\ \citenamefont {Hill}}]{WhiteHill2022}%
  \BibitemOpen
  \bibfield  {author} {\bibinfo {author} {\bibfnamefont {G.}~\bibnamefont
  {White}}, \bibinfo {author} {\bibfnamefont {F.}~\bibnamefont {Pollock}},
  \bibinfo {author} {\bibfnamefont {L.}~\bibnamefont {Hollenberg}}, \bibinfo
  {author} {\bibfnamefont {K.}~\bibnamefont {Modi}}, \ and\ \bibinfo {author}
  {\bibfnamefont {C.}~\bibnamefont {Hill}},\ }\href {\doibase
  10.1103/PRXQuantum.3.020344} {\bibfield  {journal} {\bibinfo  {journal} {PRX
  Quantum}\ }\textbf {\bibinfo {volume} {3}},\ \bibinfo {pages} {020344}
  (\bibinfo {year} {2022})}\BibitemShut {NoStop}%
\bibitem [{\citenamefont {White}\ \emph {et~al.}(2021)\citenamefont {White},
  \citenamefont {Pollock}, \citenamefont {Hollenberg}, \citenamefont {Hill},\
  and\ \citenamefont {Modi}}]{white2021many}%
  \BibitemOpen
  \bibfield  {author} {\bibinfo {author} {\bibfnamefont {G.~A.~L.}\
  \bibnamefont {White}}, \bibinfo {author} {\bibfnamefont {F.~A.}\ \bibnamefont
  {Pollock}}, \bibinfo {author} {\bibfnamefont {L.~C.~L.}\ \bibnamefont
  {Hollenberg}}, \bibinfo {author} {\bibfnamefont {C.~D.}\ \bibnamefont
  {Hill}}, \ and\ \bibinfo {author} {\bibfnamefont {K.}~\bibnamefont {Modi}},\
  }\href {http://arxiv.org/abs/2107.13934} {\bibfield  {journal} {\bibinfo
  {journal} {arXiv:2107.13934}\ } (\bibinfo {year} {2021})}\BibitemShut
  {NoStop}%
\bibitem [{\citenamefont {Luchnikov}\ \emph {et~al.}(2019)\citenamefont
  {Luchnikov}, \citenamefont {Vintskevich}, \citenamefont {Ouerdane},\ and\
  \citenamefont {Filippov}}]{LuchnikovFilippov2019}%
  \BibitemOpen
  \bibfield  {author} {\bibinfo {author} {\bibfnamefont {I.~A.}\ \bibnamefont
  {Luchnikov}}, \bibinfo {author} {\bibfnamefont {S.~V.}\ \bibnamefont
  {Vintskevich}}, \bibinfo {author} {\bibfnamefont {H.}~\bibnamefont
  {Ouerdane}}, \ and\ \bibinfo {author} {\bibfnamefont {S.~N.}\ \bibnamefont
  {Filippov}},\ }\href {\doibase 10.1103/PhysRevLett.122.160401} {\bibfield
  {journal} {\bibinfo  {journal} {Phys. Rev. Lett.}\ }\textbf {\bibinfo
  {volume} {122}},\ \bibinfo {pages} {160401} (\bibinfo {year}
  {2019})}\BibitemShut {NoStop}%
\bibitem [{\citenamefont {Guo}\ \emph {et~al.}(2018)\citenamefont {Guo},
  \citenamefont {Jie}, \citenamefont {Lu},\ and\ \citenamefont
  {Poletti}}]{GuoPoletti2018}%
  \BibitemOpen
  \bibfield  {author} {\bibinfo {author} {\bibfnamefont {C.}~\bibnamefont
  {Guo}}, \bibinfo {author} {\bibfnamefont {Z.}~\bibnamefont {Jie}}, \bibinfo
  {author} {\bibfnamefont {W.}~\bibnamefont {Lu}}, \ and\ \bibinfo {author}
  {\bibfnamefont {D.}~\bibnamefont {Poletti}},\ }\href {\doibase
  10.1103/PhysRevE.98.042114} {\bibfield  {journal} {\bibinfo  {journal} {Phys.
  Rev. E}\ }\textbf {\bibinfo {volume} {98}},\ \bibinfo {pages} {042114}
  (\bibinfo {year} {2018})}\BibitemShut {NoStop}%
\bibitem [{\citenamefont {Guo}\ \emph {et~al.}(2020)\citenamefont {Guo},
  \citenamefont {Modi},\ and\ \citenamefont {Poletti}}]{GuoPoletti2020}%
  \BibitemOpen
  \bibfield  {author} {\bibinfo {author} {\bibfnamefont {C.}~\bibnamefont
  {Guo}}, \bibinfo {author} {\bibfnamefont {K.}~\bibnamefont {Modi}}, \ and\
  \bibinfo {author} {\bibfnamefont {D.}~\bibnamefont {Poletti}},\ }\href
  {\doibase 10.1103/PhysRevA.102.062414} {\bibfield  {journal} {\bibinfo
  {journal} {Phys. Rev. A}\ }\textbf {\bibinfo {volume} {102}},\ \bibinfo
  {pages} {062414} (\bibinfo {year} {2020})}\BibitemShut {NoStop}%
\bibitem [{\citenamefont {Luchnikov}\ \emph {et~al.}(2020)\citenamefont
  {Luchnikov}, \citenamefont {Vintskevich}, \citenamefont {Grigoriev},\ and\
  \citenamefont {Filippov}}]{LuchnikovFilippov2020}%
  \BibitemOpen
  \bibfield  {author} {\bibinfo {author} {\bibfnamefont {I.~A.}\ \bibnamefont
  {Luchnikov}}, \bibinfo {author} {\bibfnamefont {S.~V.}\ \bibnamefont
  {Vintskevich}}, \bibinfo {author} {\bibfnamefont {D.~A.}\ \bibnamefont
  {Grigoriev}}, \ and\ \bibinfo {author} {\bibfnamefont {S.~N.}\ \bibnamefont
  {Filippov}},\ }\href {\doibase 10.1103/PhysRevLett.124.140502} {\bibfield
  {journal} {\bibinfo  {journal} {Phys. Rev. Lett.}\ }\textbf {\bibinfo
  {volume} {124}},\ \bibinfo {pages} {140502} (\bibinfo {year}
  {2020})}\BibitemShut {NoStop}%
\bibitem [{\citenamefont {White}\ \emph {et~al.}(2023)\citenamefont {White},
  \citenamefont {Modi},\ and\ \citenamefont {Hill}}]{PhysRevLett.130.160401}%
  \BibitemOpen
  \bibfield  {author} {\bibinfo {author} {\bibfnamefont {G.~A.~L.}\
  \bibnamefont {White}}, \bibinfo {author} {\bibfnamefont {K.}~\bibnamefont
  {Modi}}, \ and\ \bibinfo {author} {\bibfnamefont {C.~D.}\ \bibnamefont
  {Hill}},\ }\href {\doibase 10.1103/PhysRevLett.130.160401} {\bibfield
  {journal} {\bibinfo  {journal} {Phys. Rev. Lett.}\ }\textbf {\bibinfo
  {volume} {130}},\ \bibinfo {pages} {160401} (\bibinfo {year}
  {2023})}\BibitemShut {NoStop}%
\bibitem [{\citenamefont {Crutchfield}\ and\ \citenamefont
  {Young}(1989)}]{CrutchfieldYong1989}%
  \BibitemOpen
  \bibfield  {author} {\bibinfo {author} {\bibfnamefont {J.~P.}\ \bibnamefont
  {Crutchfield}}\ and\ \bibinfo {author} {\bibfnamefont {K.}~\bibnamefont
  {Young}},\ }\href {\doibase 10.1103/PhysRevLett.63.105} {\bibfield  {journal}
  {\bibinfo  {journal} {Phys. Rev. Lett.}\ }\textbf {\bibinfo {volume} {63}},\
  \bibinfo {pages} {105} (\bibinfo {year} {1989})}\BibitemShut {NoStop}%
\bibitem [{\citenamefont {Shalizi}\ and\ \citenamefont
  {Crutchfield}(2001)}]{ShaliziCrutchfield2001}%
  \BibitemOpen
  \bibfield  {author} {\bibinfo {author} {\bibfnamefont {C.~R.}\ \bibnamefont
  {Shalizi}}\ and\ \bibinfo {author} {\bibfnamefont {J.~P.}\ \bibnamefont
  {Crutchfield}},\ }\href {\doibase 10.1023/A:1010388907793} {\bibfield
  {journal} {\bibinfo  {journal} {Journal of Statistical Physics}\ }\textbf
  {\bibinfo {volume} {104}},\ \bibinfo {pages} {817} (\bibinfo {year}
  {2001})}\BibitemShut {NoStop}%
\bibitem [{\citenamefont {Guo}(2022{\natexlab{a}})}]{Guo2022a}%
  \BibitemOpen
  \bibfield  {author} {\bibinfo {author} {\bibfnamefont {C.}~\bibnamefont
  {Guo}},\ }\href {\doibase 10.1103/PhysRevA.106.022411} {\bibfield  {journal}
  {\bibinfo  {journal} {Phys. Rev. A}\ }\textbf {\bibinfo {volume} {106}},\
  \bibinfo {pages} {022411} (\bibinfo {year} {2022}{\natexlab{a}})}\BibitemShut
  {NoStop}%
\bibitem [{sup()}]{supp}%
  \BibitemOpen
  \href@noop {} {\enquote {\bibinfo {title} {See supplementary material.}}\
  }\BibitemShut {NoStop}%
\bibitem [{\citenamefont {Figueroa-Romero}\ \emph {et~al.}(2021)\citenamefont
  {Figueroa-Romero}, \citenamefont {Modi}, \citenamefont {Harris},
  \citenamefont {Stace},\ and\ \citenamefont {Hsieh}}]{RomeroHsieh2021}%
  \BibitemOpen
  \bibfield  {author} {\bibinfo {author} {\bibfnamefont {P.}~\bibnamefont
  {Figueroa-Romero}}, \bibinfo {author} {\bibfnamefont {K.}~\bibnamefont
  {Modi}}, \bibinfo {author} {\bibfnamefont {R.~J.}\ \bibnamefont {Harris}},
  \bibinfo {author} {\bibfnamefont {T.~M.}\ \bibnamefont {Stace}}, \ and\
  \bibinfo {author} {\bibfnamefont {M.-H.}\ \bibnamefont {Hsieh}},\ }\href
  {\doibase 10.1103/PRXQuantum.2.040351} {\bibfield  {journal} {\bibinfo
  {journal} {PRX Quantum}\ }\textbf {\bibinfo {volume} {2}},\ \bibinfo {pages}
  {040351} (\bibinfo {year} {2021})}\BibitemShut {NoStop}%
\bibitem [{\citenamefont {Fletcher}(2000)}]{BFGS}%
  \BibitemOpen
  \bibfield  {author} {\bibinfo {author} {\bibfnamefont {R.}~\bibnamefont
  {Fletcher}},\ }\enquote {\bibinfo {title} {Newton-like methods},}\ in\ \href
  {\doibase https://doi.org/10.1002/9781118723203.ch3} {\emph {\bibinfo
  {booktitle} {Practical Methods of Optimization}}}\ (\bibinfo  {publisher}
  {John Wiley \& Sons, Ltd},\ \bibinfo {year} {2000})\ Chap.~\bibinfo {chapter}
  {3}, pp.\ \bibinfo {pages} {44--79}\BibitemShut {NoStop}%
\bibitem [{\citenamefont {Reck}\ \emph {et~al.}(1994)\citenamefont {Reck},
  \citenamefont {Zeilinger}, \citenamefont {Bernstein},\ and\ \citenamefont
  {Bertani}}]{ReckBertani1994}%
  \BibitemOpen
  \bibfield  {author} {\bibinfo {author} {\bibfnamefont {M.}~\bibnamefont
  {Reck}}, \bibinfo {author} {\bibfnamefont {A.}~\bibnamefont {Zeilinger}},
  \bibinfo {author} {\bibfnamefont {H.~J.}\ \bibnamefont {Bernstein}}, \ and\
  \bibinfo {author} {\bibfnamefont {P.}~\bibnamefont {Bertani}},\ }\href
  {\doibase 10.1103/PhysRevLett.73.58} {\bibfield  {journal} {\bibinfo
  {journal} {Phys. Rev. Lett.}\ }\textbf {\bibinfo {volume} {73}},\ \bibinfo
  {pages} {58} (\bibinfo {year} {1994})}\BibitemShut {NoStop}%
\bibitem [{\citenamefont {Guo}\ and\ \citenamefont
  {Poletti}(2021)}]{GuoPoletti2021}%
  \BibitemOpen
  \bibfield  {author} {\bibinfo {author} {\bibfnamefont {C.}~\bibnamefont
  {Guo}}\ and\ \bibinfo {author} {\bibfnamefont {D.}~\bibnamefont {Poletti}},\
  }\href {\doibase 10.1103/PhysRevE.103.013309} {\bibfield  {journal} {\bibinfo
   {journal} {Phys. Rev. E}\ }\textbf {\bibinfo {volume} {103}},\ \bibinfo
  {pages} {013309} (\bibinfo {year} {2021})}\BibitemShut {NoStop}%
\bibitem [{\citenamefont {Epstein}\ \emph {et~al.}(2014)\citenamefont
  {Epstein}, \citenamefont {Cross}, \citenamefont {Magesan},\ and\
  \citenamefont {Gambetta}}]{EpsteinGambetta2014}%
  \BibitemOpen
  \bibfield  {author} {\bibinfo {author} {\bibfnamefont {J.~M.}\ \bibnamefont
  {Epstein}}, \bibinfo {author} {\bibfnamefont {A.~W.}\ \bibnamefont {Cross}},
  \bibinfo {author} {\bibfnamefont {E.}~\bibnamefont {Magesan}}, \ and\
  \bibinfo {author} {\bibfnamefont {J.~M.}\ \bibnamefont {Gambetta}},\ }\href
  {\doibase 10.1103/PhysRevA.89.062321} {\bibfield  {journal} {\bibinfo
  {journal} {Phys. Rev. A}\ }\textbf {\bibinfo {volume} {89}},\ \bibinfo
  {pages} {062321} (\bibinfo {year} {2014})}\BibitemShut {NoStop}%
\bibitem [{\citenamefont {Yang}\ \emph {et~al.}(2018)\citenamefont {Yang},
  \citenamefont {Binder}, \citenamefont {Narasimhachar},\ and\ \citenamefont
  {Gu}}]{YangGu2018}%
  \BibitemOpen
  \bibfield  {author} {\bibinfo {author} {\bibfnamefont {C.}~\bibnamefont
  {Yang}}, \bibinfo {author} {\bibfnamefont {F.~C.}\ \bibnamefont {Binder}},
  \bibinfo {author} {\bibfnamefont {V.}~\bibnamefont {Narasimhachar}}, \ and\
  \bibinfo {author} {\bibfnamefont {M.}~\bibnamefont {Gu}},\ }\href {\doibase
  10.1103/PhysRevLett.121.260602} {\bibfield  {journal} {\bibinfo  {journal}
  {Phys. Rev. Lett.}\ }\textbf {\bibinfo {volume} {121}},\ \bibinfo {pages}
  {260602} (\bibinfo {year} {2018})}\BibitemShut {NoStop}%
\bibitem [{\citenamefont {Guo}(2022{\natexlab{b}})}]{Guo2022c}%
  \BibitemOpen
  \bibfield  {author} {\bibinfo {author} {\bibfnamefont {C.}~\bibnamefont
  {Guo}},\ }\href {\doibase 10.21468/SciPostPhys.13.2.028} {\bibfield
  {journal} {\bibinfo  {journal} {SciPost Phys.}\ }\textbf {\bibinfo {volume}
  {13}},\ \bibinfo {pages} {028} (\bibinfo {year}
  {2022}{\natexlab{b}})}\BibitemShut {NoStop}%
\end{thebibliography}%


\begin{thebibliography}{6}%
\makeatletter
\providecommand \@ifxundefined [1]{%
 \@ifx{#1\undefined}
}%
\providecommand \@ifnum [1]{%
 \ifnum #1\expandafter \@firstoftwo
 \else \expandafter \@secondoftwo
 \fi
}%
\providecommand \@ifx [1]{%
 \ifx #1\expandafter \@firstoftwo
 \else \expandafter \@secondoftwo
 \fi
}%
\providecommand \natexlab [1]{#1}%
\providecommand \enquote  [1]{``#1''}%
\providecommand \bibnamefont  [1]{#1}%
\providecommand \bibfnamefont [1]{#1}%
\providecommand \citenamefont [1]{#1}%
\providecommand \href@noop [0]{\@secondoftwo}%
\providecommand \href [0]{\begingroup \@sanitize@url \@href}%
\providecommand \@href[1]{\@@startlink{#1}\@@href}%
\providecommand \@@href[1]{\endgroup#1\@@endlink}%
\providecommand \@sanitize@url [0]{\catcode `\\12\catcode `\$12\catcode
  `\&12\catcode `\#12\catcode `\^12\catcode `\_12\catcode `\%12\relax}%
\providecommand \@@startlink[1]{}%
\providecommand \@@endlink[0]{}%
\providecommand \url  [0]{\begingroup\@sanitize@url \@url }%
\providecommand \@url [1]{\endgroup\@href {#1}{\urlprefix }}%
\providecommand \urlprefix  [0]{URL }%
\providecommand \Eprint [0]{\href }%
\providecommand \doibase [0]{https://doi.org/}%
\providecommand \selectlanguage [0]{\@gobble}%
\providecommand \bibinfo  [0]{\@secondoftwo}%
\providecommand \bibfield  [0]{\@secondoftwo}%
\providecommand \translation [1]{[#1]}%
\providecommand \BibitemOpen [0]{}%
\providecommand \bibitemStop [0]{}%
\providecommand \bibitemNoStop [0]{.\EOS\space}%
\providecommand \EOS [0]{\spacefactor3000\relax}%
\providecommand \BibitemShut  [1]{\csname bibitem#1\endcsname}%
\let\auto@bib@innerbib\@empty
\bibitem [{\citenamefont {Epstein}\ \emph {et~al.}(2014)\citenamefont
  {Epstein}, \citenamefont {Cross}, \citenamefont {Magesan},\ and\
  \citenamefont {Gambetta}}]{EpsteinGambetta2014}%
  \BibitemOpen
  \bibfield  {author} {\bibinfo {author} {\bibfnamefont {J.~M.}\ \bibnamefont
  {Epstein}}, \bibinfo {author} {\bibfnamefont {A.~W.}\ \bibnamefont {Cross}},
  \bibinfo {author} {\bibfnamefont {E.}~\bibnamefont {Magesan}},\ and\ \bibinfo
  {author} {\bibfnamefont {J.~M.}\ \bibnamefont {Gambetta}},\ }\href
  {https://doi.org/10.1103/PhysRevA.89.062321} {\bibfield  {journal} {\bibinfo
  {journal} {Phys. Rev. A}\ }\textbf {\bibinfo {volume} {89}},\ \bibinfo
  {pages} {062321} (\bibinfo {year} {2014})}\BibitemShut {NoStop}%
\bibitem [{\citenamefont {Verstraete}\ \emph {et~al.}(2004)\citenamefont
  {Verstraete}, \citenamefont {Garc\'{\i}a-Ripoll},\ and\ \citenamefont
  {Cirac}}]{VerstraeteCirac2004}%
  \BibitemOpen
  \bibfield  {author} {\bibinfo {author} {\bibfnamefont {F.}~\bibnamefont
  {Verstraete}}, \bibinfo {author} {\bibfnamefont {J.~J.}\ \bibnamefont
  {Garc\'{\i}a-Ripoll}},\ and\ \bibinfo {author} {\bibfnamefont {J.~I.}\
  \bibnamefont {Cirac}},\ }\href
  {https://doi.org/10.1103/PhysRevLett.93.207204} {\bibfield  {journal}
  {\bibinfo  {journal} {Phys. Rev. Lett.}\ }\textbf {\bibinfo {volume} {93}},\
  \bibinfo {pages} {207204} (\bibinfo {year} {2004})}\BibitemShut {NoStop}%
\bibitem [{\citenamefont {Pollock}\ \emph {et~al.}(2018)\citenamefont
  {Pollock}, \citenamefont {Rodr\'{\i}guez-Rosario}, \citenamefont
  {Frauenheim}, \citenamefont {Paternostro},\ and\ \citenamefont
  {Modi}}]{PollockModi2018b}%
  \BibitemOpen
  \bibfield  {author} {\bibinfo {author} {\bibfnamefont {F.~A.}\ \bibnamefont
  {Pollock}}, \bibinfo {author} {\bibfnamefont {C.}~\bibnamefont
  {Rodr\'{\i}guez-Rosario}}, \bibinfo {author} {\bibfnamefont {T.}~\bibnamefont
  {Frauenheim}}, \bibinfo {author} {\bibfnamefont {M.}~\bibnamefont
  {Paternostro}},\ and\ \bibinfo {author} {\bibfnamefont {K.}~\bibnamefont
  {Modi}},\ }\href {https://doi.org/10.1103/PhysRevLett.120.040405} {\bibfield
  {journal} {\bibinfo  {journal} {Phys. Rev. Lett.}\ }\textbf {\bibinfo
  {volume} {120}},\ \bibinfo {pages} {040405} (\bibinfo {year}
  {2018})}\BibitemShut {NoStop}%
\bibitem [{\citenamefont {Schollwöck}(2011)}]{Schollwock2011}%
  \BibitemOpen
  \bibfield  {author} {\bibinfo {author} {\bibfnamefont {U.}~\bibnamefont
  {Schollwöck}},\ }\href
  {https://doi.org/https://doi.org/10.1016/j.aop.2010.09.012} {\bibfield
  {journal} {\bibinfo  {journal} {Annals of Physics}\ }\textbf {\bibinfo
  {volume} {326}},\ \bibinfo {pages} {96} (\bibinfo {year} {2011})}\BibitemShut
  {NoStop}%
\bibitem [{\citenamefont {Pi{\v{z}}orn}\ and\ \citenamefont
  {Prosen}(2009)}]{IztokProsen2009}%
  \BibitemOpen
  \bibfield  {author} {\bibinfo {author} {\bibfnamefont {I.}~\bibnamefont
  {Pi{\v{z}}orn}}\ and\ \bibinfo {author} {\bibfnamefont {T.}~\bibnamefont
  {Prosen}},\ }\href {https://doi.org/10.1103/PhysRevB.79.184416} {\bibfield
  {journal} {\bibinfo  {journal} {Phys. Rev. B}\ }\textbf {\bibinfo {volume}
  {79}},\ \bibinfo {pages} {184416} (\bibinfo {year} {2009})}\BibitemShut
  {NoStop}%
\bibitem [{\citenamefont {Guo}(2022)}]{Guo2022c}%
  \BibitemOpen
  \bibfield  {author} {\bibinfo {author} {\bibfnamefont {C.}~\bibnamefont
  {Guo}},\ }\href {https://doi.org/10.21468/SciPostPhys.13.2.028} {\bibfield
  {journal} {\bibinfo  {journal} {SciPost Phys.}\ }\textbf {\bibinfo {volume}
  {13}},\ \bibinfo {pages} {028} (\bibinfo {year} {2022})}\BibitemShut
  {NoStop}%
\end{thebibliography}%

\end{document}


\title{Supplementary Information: Randomised benchmarking for characterizing and forecasting correlated processes}

\author{Xinfang Zhang}
\thanks{These two authors contributed equally}
\affiliation{\nudt}

\author{Zhihao Wu}
\thanks{These two authors contributed equally}
\affiliation{\nudt}

\author{Gregory A. L. White}
\affiliation{School of Physics and Astronomy, Monash University, Victoria 3800, Australia}
\affiliation{Dahlem Center for Complex Quantum Systems, Freie Universit\"at Berlin, 14195 Berlin, Germany}

\author{Zhongcheng Xiang}
\affiliation{\cas}

\author{Shun Hu}
\affiliation{\nudt}

\author{Zhihui Peng}
\affiliation{\hnu}

\author{Yong Liu}
\affiliation{\nudt}

\author{Dongning Zheng}
\affiliation{\cas}

\author{Xiang Fu}
\affiliation{\nudt}

\author{Anqi Huang}
\affiliation{\nudt}

\author{Dario Poletti} 
\email{dario\_poletti@sutd.edu.sg}
\affiliation{Science, Mathematics and Technology Cluster and Engineering Product Development Pillar, Singapore University of Technology and Design, 8 Somapah Road, 487372 Singapore} 
\affiliation{\cqt} 
\affiliation{\majulab}

\author{Kavan Modi}
\email{kavan.modi@monash.edu}
\affiliation{School of Physics and Astronomy, Monash University, Victoria 3800, Australia}
\affiliation{Quantum for NSW, Sydney 2000, Australia}

\author{Junjie Wu}
\affiliation{\nudt}

\author{Mingtang Deng}
\email{mtdeng@nudt.edu.cn}
\affiliation{\nudt}

\author{Chu Guo}
\email{guochu604b@gmail.com}
\affiliation{\hnu}

\date{\today}

\maketitle

\setcounter{figure}{0} 
\renewcommand{\thefigure}{S\arabic{figure}}
\renewcommand{\theequation}{S\arabic{equation}}
\renewcommand{\thesection}{S.\Roman{section}}

\section{Device setup}\label{sec:SectionI}

\begin{figure}
\includegraphics[width=\columnwidth]{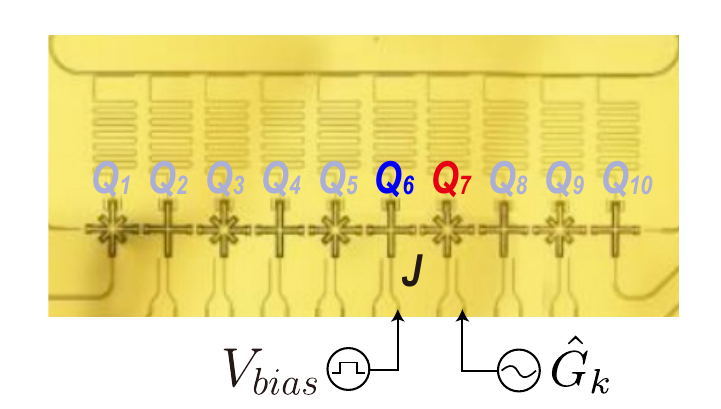}
\caption{A photograph of the superconducting quantum processor used in this work, where the two qubits $Q_7$ and $Q_6$ are used as the system ($S$) and environment ($E$) qubits respectively. The voltage bias $V_{bias}$ is applied on the environment qubit to tune the effective coupling strength between $S$ and $E$. The gate operations $\Gop_k$ are applied on $S$ for the randomised benchmarking experiment.} \label{fig:figS1}
\end{figure}

The chip we have used is shown in Fig.~\ref{fig:figS1}, which contains $10$ transmon qubits (labeled by $Q_j$ with $1\leq j\leq 10$) and $10$ dedicated resonators arranged in a 1D configuration.
Each qubit is coupled to a dedicated resonator, enabling independent readout via a common feedline. The qubit transition frequencies are controlled individually through flux bias lines, which are short-circuited near the corresponding transmon SQUID loops. The qubits are also capacitively coupled to their nearest neighbors. The total Hamiltonian can be written as
\begin{align}
\Hop=\sum_{j=1}^{10} h_{j}\sgz_{j} +\sum_{j=1}^{9} g_{j,j+1}\left ( \sgp_{j} \sgm_{j+1} + \sgm_{j}\sgp_{j+1}\right ).
\label{eq:Hamiltonian}
\end{align}

\begin{table}[t!]
\caption{Basic parameters for $Q_{S}$ and $Q_{E}$, which include the readout resonator frequency $\omega_c$, the transition frequency $\omega_{01}$ from $\ket{0}$ to $\ket{1}$, the anharmonicity $\eta$, the energy decay time $T_1$, the Ramsey decay time $T_2^*$, and the coupling strength $J$ between $S$ and $E$ (we set $\hbar=1$).}
\label{table-1}
\begin{tabular}{@{}l|c|c@{}}
\toprule  
\hline
\hline
Qubit     & $Q_{E}$& $Q_{S}$ \\ \midrule
\hline
\hline
$\omega_c /2\pi$(GHz)   & 6.6185 & 6.6384 \\
$\omega_{01}2\pi$ (GHz) & 5.5728 & 5.1543\\
$\eta /2\pi$ (MHz)      & -249   & -204 \\
$T_1$ ($\mu$ s)         & 38      & 52\\
$T_2^*$ ($\mu$ s)       & 13      & 36\\
\hline
$J$ (MHz)               & \multicolumn{2}{c}{11.30}  \\ 
\hline
\end{tabular}
\label{table:Cells}
\end{table}

In our experiment, we focus on the two qubits $Q_7$ and $Q_6$, where we take the first as system ($S$) and the second as environment ($E$).  
The characterizations of qubits $E$ and $S$ are summarized in Table \ref{table:Cells}. 
In our setup the coupling strength $g_{j, j+1}$ is fixed and we denote $J = g_{S,E} \approx 11.3$ MHz. To tune the effective coupling between $S$ and $E$, we change their imbalance $\imbalance = h_S - h_E$ by applying a voltage bias $V_{bias}$ on qubit $E$ (The voltage bias on $S$ is fixed to allow gate operations on $S$), which results in the effective coupling strength $ \gamma_{\rm eff} = 2J^2/ \imbalance$. 
Meanwhile, the frequencies of the rest qubits are tuned far away (their imbalances with respect to $S$ and $E$ are larger than $500$ MHz), such that two qubits $S$ and $E$ can be considered as isolated. 
The effective coupling strength $ \gamma_{\rm eff}$ as a function of $V_{bias}$ is shown in Fig. 2(a) of the main text. 
For $\gamma_{\rm eff}/J \ll 1$, $E$ and $S$ are effectively decoupled, while $E$ and $S$ are strongly coupled for $\gamma_{\rm eff}/J \approx 1$. Thus, we can manipulate $V_{bias}$ to tune the system dynamics from Markovian regime to highly non-Markovian regime.

\begin{figure*}[t!]
\includegraphics[width=0.98\textwidth]{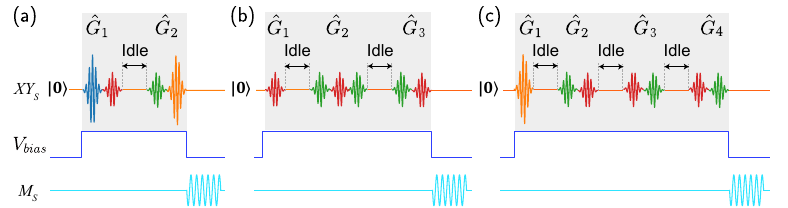}
\caption{
Microwave sequence $XY_S$ (for native gates in Eq.(\ref{eq:native})), voltage bias $V_{bias}$, and measurement pulses $M_S$ for the randomised benchmarking experiment with (a) $k=2$, (b) $k=3$ and (c) $k=4$ respectively. The system qubit $S$ is initialized to $\ket{0^S}$ state and then operated on by the native gates. The pulse duration of each native gate is fixed at $20$ ns and the idle time between the two Clifford gates is $100$ ns. The voltage bias $V_{bias}$ is applied throughout the Clifford gate operations, and ends before the final measurement pulse. }  \label{fig:figS2} 
\end{figure*}

\section{Randomised benchmarking experiment}\label{sec:SectionII}

We employ the Clifford-based randomised benchmarking (RB) to generate our training and testing data. 
The single-qubit Clifford group is the group of 24 rotations that preserve the octahedron on the Bloch sphere~\cite{EpsteinGambetta2014}. Each Clifford gate can be decomposed into rotations around the X and Y axis using the generators (native gates):
\begin{align}\label{eq:native}
\left \{\pm {\rm X/2, \pm Y/2, \pm X, \pm Y, I}\right \}.
\end{align}
We prepare RB datasets with different effective coupling strengths $\gamma_{\rm eff}$. For each $\gamma_{\rm eff}$, the detailed experimental procedures for a specific value of $k$ are as follows:

1. Initialize $S$ to its ground state $\ket{0^S}$.

2. Apply a voltage bias $V_{bias}$ on $E$, and perform random Clifford gates $\Gop_{k} \cdots \Gop_2 \Gop_1$  on qubit $S$, where each Clifford gate $\Gop_j$ is decomposed into several native gates. The idle time between each Clifford gates is set to be $100$ ns. 

3. Measure the expectation value of the $\ket{0^S}$ state of $S$.

4. Repeat the above steps for $n=200$ times, where the random Clifford gates are not the same each time.

The pulse sequence used in the experiment is shown in Fig.~\ref{fig:figS2}, exemplified for the cases $2\leq k\leq 4$.

\section{Calculating the process tensor from the open quantum evolution model}\label{sec:pt}

\begin{figure}
\includegraphics[width=\columnwidth]{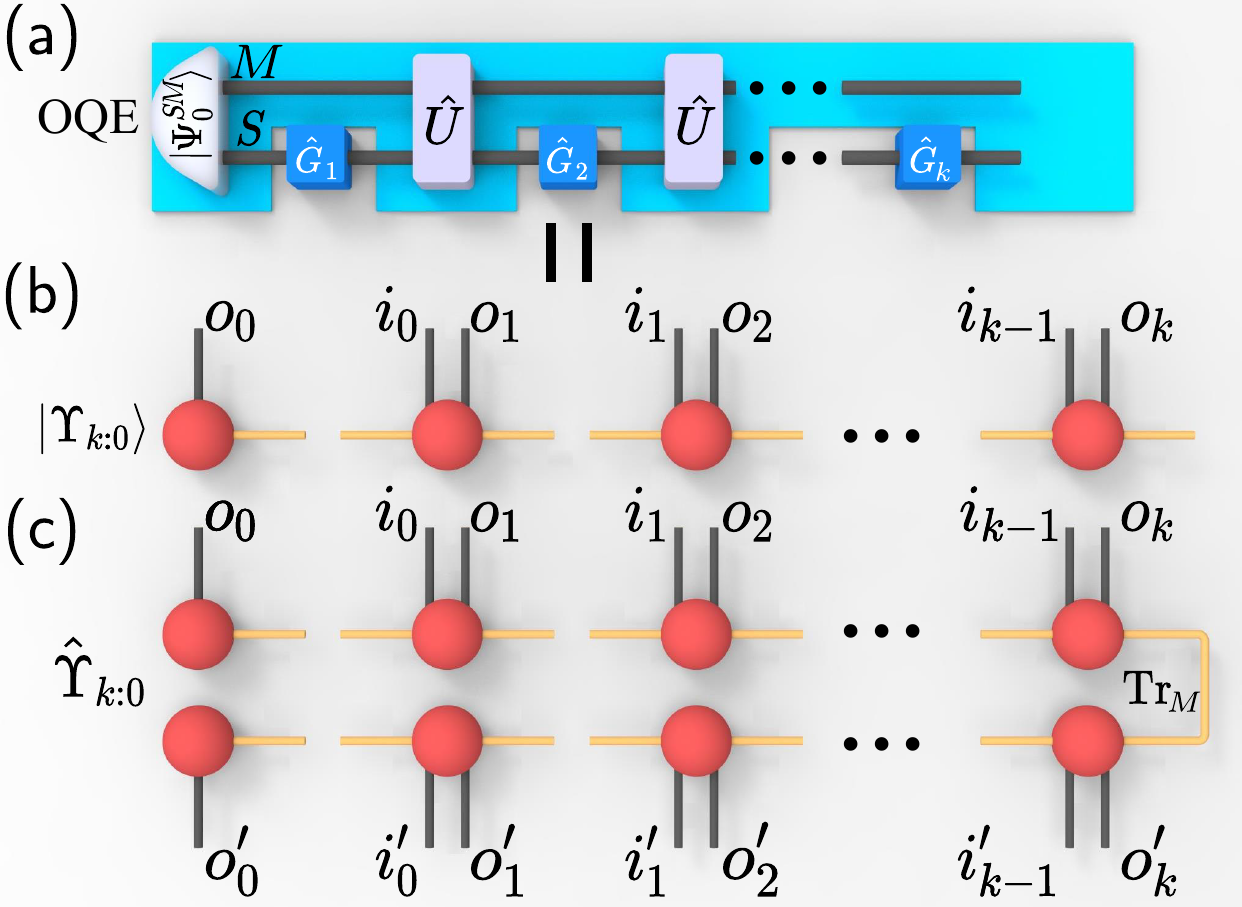}
\caption{(a) The open quantum evolution model with time-independent system-memory evolution $\Uop$. (b) The purified process tensor $\vert \Upsilon_{k:0}\rangle$ corresponding to the shaded regime in (a), which can be written as an MPS in Eq.(\ref{eq:ppt}). (c) The process tensor $\hat{\Upsilon}_{k:0}$ is an MPDO obtained by tracing out the environment degrees of freedom in the purified process tensor.
}   \label{fig:oqe} 
\end{figure}

The multi-time quantum dynamics described by the open quantum evolution (OQE) model, as shown in Fig.~\ref{fig:oqe}(a), 
can be equivalently represented as a matrix product state (MPS):
\begin{align}\label{eq:ppt}
\vert \Upsilon_{k:0}\rangle = \sum_{\bm{o}, \bm{i}, \bm{\alpha}} B^{o_0}_{\alpha_0}B^{i_0,o_1}_{\alpha_0,\alpha_1} \cdots B^{i_{k-1}, o_k}_{\alpha_{k-1},\alpha_k}\vert \alpha_k\rangle \vert \bm{o}, \bm{i} \rangle,
\end{align}
where the site tensors are related to the OQE as
\begin{align}
B^{o_0}_{\alpha_0} &= \langle o_0, \alpha_0\vert \Psi_0^{SM}\rangle; \\
B^{i_{j-1}, o_j}_{\alpha_{j-1}, \alpha_j} &= \langle i_{j-1}, \alpha_{j-1} \vert \Uop_{j:j-1} \vert o_j, \alpha_j\rangle, \quad \forall 1\leq j\leq k.
\end{align}
$\vert \Upsilon_{k:0}\rangle$ is referred to as the purified process tensor (PPT), which encodes all the information about the system-memory dynamics and is shown in Fig.~\ref{fig:oqe}(b). However, as can be seen from Eq.(\ref{eq:ppt}), $\vert \Upsilon_{k:0}\rangle$ directly contains the environment degrees of freedom, and the choice of environment is not unique (for example, one can perform an arbitrary unitary transformation on the environment, which will not affect the observables on the system). To eliminate this uncertainty, one can trace out the memory index from the PPT, which results in the process tensor
\begin{align}\label{eq:pt}
\hat{\Upsilon}_{k:0} =& \Tr_M\left(\vert \Upsilon_{k:0}\rangle \langle \Upsilon_{k:0}\vert \right) \nonumber \\ 
=& \sum_{\bm{o}, \bm{o}', \bm{i}, \bm{i}', \bm{\alpha}, \bm{\alpha}'} W^{o_0, o_0'}_{\alpha_0, \alpha_0'} W^{i_0, i_0', o_1, o_1'}_{\alpha_0, \alpha_0', \alpha_1, \alpha_1'} \times \cdots \nonumber \\
&\times W^{i_{k-1}, i_{k-1}', o_k, o_k'}_{\alpha_{k-1}, \alpha_{k-1}'}  \vert \bm{o}, \bm{i} \rangle\langle \bm{o}', \bm{i}'\vert .
\end{align}
Here the site tensors are
\begin{subequations}\label{eq:Wtensors}
\begin{align}
W^{o_0, o_0'}_{\alpha_0, \alpha_0'} &= B^{o_0}_{\alpha_0} B^{o_0'}_{\alpha_0'}; \\
W^{i_{j-1}, i_{j-1}', o_j, o_j'}_{\alpha_{j-1}, \alpha_{j-1}', \alpha_j, \alpha_j'} &=  B^{i_{j-1}, o_j}_{\alpha_{j-1},\alpha_j} B^{i_{j-1}', o_j'}_{\alpha_{j-1}',\alpha_j'}, \quad \forall 1\leq j < k ; \\
W^{i_{k-1}, i_{k-1}', o_k, o_k'}_{\alpha_{k-1}, \alpha_{k-1}'} &= \sum_{\alpha_k} B^{i_{k-1}, o_k}_{\alpha_{k-1},\alpha_k} B^{i_{k-1}', o_k'}_{\alpha_{k-1}',\alpha_k}.
\end{align}
\end{subequations}
The process tensor only contains the system degrees of freedom and is thus unique, as can be seen from Eq.(\ref{eq:pt}).
Moreover, from Eqs.(\ref{eq:Wtensors}) we can see that each site tensor of the process tensor is positive semi-definite, hence the process tensor $\hat{\Upsilon}_{k:0}$ defined in Eq.(\ref{eq:pt}) is naturally a matrix product density operator (MPDO)~\cite{VerstraeteCirac2004}.

\section{Calculating the non-Markovianity measures}
In this section we describe the details of calculating different non-Markovianity measures used in the main text.
Since the process tensor contains all the observable information, any non-Markovianity measure should only depend on the process tensor in principle.

An operational condition for non-Markovianity can be determined by considering its closest Markovian process tensor under the relative entropy quasi-distance, which is the product state 
\begin{align}
\Upsilon_{k:0}^{\text{Markov}} &:= \hat{\mathcal{E}}_{k:k-1}\otimes \hat{\mathcal{E}}_{k-1:k-2}\otimes \cdots \otimes \hat{\mathcal{E}}_{1:0},
\end{align}
where
\begin{align}\label{eq:pt2}
\hat{\mathcal{E}}_{j:j-1} = \Trp_{\overline{o_ji_j}}[\Upsilon_{k:0}],
\end{align}
with $\overline{o_ji_j}$ denoting every index of the tensor except $o_j$ and $i_j$. 
We have also defined a modified trace operation $\Trp$, which means that 
for the indices that are traced out, $i_\ell$ is contracted with $o_{\ell-1}$ and $i_\ell'$ is contracted with $o_{\ell-1}'$ (instead of the normal trace where $i_\ell$ is contracted with $i_\ell'$). This is equivalent to propagating the system with an identity operation at each $t_\ell$.
The quantity 
\begin{align} \label{eq:nm}
    \mathcal{N}(\Upsilon_{k:0}) &:= \mathcal{S}\left(\Upsilon_{k:0}\mid\mid \Upsilon_{k:0}^{\text{Markov}}\right) \nonumber \\
    &= \text{Tr}[\Upsilon_{k:0}(\log \Upsilon_{k:0} - \log \Upsilon_{k:0}^{\text{Markov}})]
\end{align}
is a measure of non-Markovianity, as introduced in Ref.~\cite{PollockModi2018b}. This can be interpreted roughly as the probability of confusing the Markov model for the true process given $\lambda$ experiments,
\begin{align}
    \mathbb{P}(\text{accept } \Upsilon_{k:0}^{\text{Markov}}) = \exp\left(-\lambda \mathcal{N}(\Upsilon_{k:0})\right).
\end{align}
In full generality, this property becomes exponentially difficult to compute with $k$. Moreover, this single figure is not indicative of how memory might dynamically change throughout the process. 
In the main text we have computed three versions of non-Markovianity measures, which can be efficiently calculated and reveal different aspects of non-Markovianity.

The memory complexity $\mathcal{M}_j$ is defined as the Von Neumann entropy of the process tensor $\hat{\Upsilon}_{j:0}$. $\hat{\Upsilon}_{j:0}$ has the same form as a many-body density operator. However, calculating the entropy of a many-body density operator written as an MPDO is a numerically hard problem. Nevertheless, we notice that the entropy of $\hat{\Upsilon}_{j:0}$ is equal to the bipartition entanglement entropy of the PPT $\hat{\Upsilon}_{k:0}$ (with $k\geq j$) at the $j$-th step which can be efficient calculated (calculating the bipartition entanglement entropy of an MPS is a standard technique whose computational cost scales as $\chi^3$~\cite{Schollwock2011}, where $\chi$ the bond dimension of MPS, which is equal to the memory size). Since we can easily obtain the PPT for any $k$ based on the reconstructed OQE, we can compute $\mathcal{M}_j$ efficiently for any $j$. 

The non-Markovianity $\mathcal{N}_j$, defined as the operator space entanglement entropy~\cite{IztokProsen2009} of the process tensor $\hat{\Upsilon}_{k:0}$ (with $k\geq j$), can be also be efficiently calculated by vectorizing $\hat{\Upsilon}_{k:0}$ into an MPS and then compute the bipartition entanglement entropy of the resulting MPS at step $j$. There is a boundary effect for this quantity for $j$ close to $k$~\cite{Guo2022c}, however, this boundary effect can be easily removed since we can directly obtain $\hat{\Upsilon}_{k:0}$ for any $k$ with the reconstructed OQE.

The mutual information $\mathcal{I}_{x, y}$ is a marginal version of Eq.(\ref{eq:nm}) that can be efficiently computed. 
From $\Upsilon_{k:0}$ we can construct a conditional process tensor $\hat{\mathcal{E}}_{x:x-1;y:y-1}$ -- which we call $\Upsilon_{x,y}$ for brevity -- which encodes temporal correlations between the dynamical maps from time $t_{y-1}$ to $t_{y}$ and from $t_{x-1}$ to $t_x$ and can be computed in accordance with Eq.(\ref{eq:pt2}).
The mutual information can then be computed as
\begin{align}
\mathcal{I}(x,y) = \mathcal{S}(\hat{\mathcal{E}}_{x:x-1;y:y-1} \mid\mid \hat{\mathcal{E}}_{x:x-1}\otimes\hat{\mathcal{E}}_{y:y-1})
\end{align}
and can be interpreted as per the above discussion in the context of two dynamical maps spanning four times.

\bibliographystyle{apsrev4-2}
\bibliography{refs_v1}